\documentclass{article} 
\usepackage{iclr2026_conference,times}
\usepackage{multirow}
\usepackage[table]{xcolor}
\definecolor{mygreen}{HTML}{E6F8E0}
\definecolor{myblue}{HTML}{CFDFF5}
\definecolor{myred}{HTML}{FFCCCC}
\definecolor{mygray}{HTML}{D3D3D3}
\usepackage{pifont}
\usepackage{booktabs}
\usepackage{arydshln}
\usepackage{subfigure}
\usepackage{amssymb}
\usepackage{tikz}
\usepackage{xcolor} 
\definecolor{mblue}{RGB}{0, 77, 128}
\definecolor{mred}{RGB}{192,0, 0}
\newcommand{\blue}[1]{\textbf{\textcolor{mblue}{#1}}}
\newcommand{\red}[1]{\textbf{\textcolor{mred}{#1}}}
\usepackage{graphicx}  
\usepackage{anyfontsize}
\newcommand{\minitiny}{\fontsize{4pt}{4.7pt}\selectfont}
\usepackage{siunitx}
\usepackage{wrapfig}
\usepackage{lipsum}

\usepackage{amsmath,amsfonts,bm}

\def\eqref#1{equation~\ref{#1}}

\def\1{\bm{1}}

\DeclareMathAlphabet{\mathsfit}{\encodingdefault}{\sfdefault}{m}{sl}
\SetMathAlphabet{\mathsfit}{bold}{\encodingdefault}{\sfdefault}{bx}{n}

\iclrfinalcopy
\usepackage{hyperref}
\usepackage{url}

\title{XGC-AVis: Towards Audio-Visual Content \\ Understanding with a Multi-Agent \\Collaborative System}

\author{Yuqin Cao$^1$, Xiongkuo Min$^1$, Yixuan Gao$^1$, Wei Sun$^2$, \\
\textbf{Zicheng Zhang}$^{1,3}$, \textbf{Jinliang Han}$^1$, \textbf{Guangtao Zhai}$^1$\\
$^1$Shanghai Jiao Tong University $^2$East China Normal University $^3$Shanghai AI Laboratory
}

\begin{document}

\maketitle

\begin{abstract} 
In this paper, we propose \textbf{XGC-AVis}, a multi-agent framework that enhances the audio-video temporal alignment capabilities of multimodal large models (MLLMs) and improves the efficiency of retrieving key video segments through $4$ stages: perception, planning, execution, and reflection. We further introduce \textbf{XGC-AVQuiz}, the first benchmark aimed at comprehensively assessing MLLMs' understanding capabilities in both real-world and AI-generated scenarios. XGC-AVQuiz consists of $2,685$ question-answer pairs across $20$ tasks, with two key innovations: 1) \textbf{AIGC Scenario Expansion:} The benchmark includes $2,232$ videos, comprising $1,102$ professionally generated content (PGC), $753$ user-generated content (UGC), and $377$ AI-generated content (AIGC). These videos cover $10$ major domains and $53$ fine-grained categories. 2) \textbf{Quality Perception Dimension:} Beyond conventional tasks such as recognition, localization, and reasoning, we introduce a novel quality perception dimension. This requires MLLMs to integrate low-level sensory capabilities with high-level semantic understanding to assess audio-visual quality, synchronization, and coherence. Experimental results on XGC-AVQuiz demonstrate that current MLLMs struggle with quality perception and temporal alignment tasks. XGC-AVis improves these capabilities without requiring additional training, as validated on two benchmarks. The project page is available at: \url{https://xgc-avis.github.io/XGC-AVis/}
\end{abstract}

\section{Introduction}
Vision and hearing play crucial roles in human perception and understanding. The human brain can simultaneously perceive multiple modalities, such as text, vision, and audio, and integrate them for joint perception and reasoning. Compared to unimodal inputs, multimodal information enables a more comprehensive understanding and reasoning through cross-modal complementarity. Consequently, enabling multimodal large language models (MLLMs) to efficiently integrate and comprehend multimodal information has become a key direction for future development.

In recent years, MLLMs \cite{li2024llava,chen2024expanding,zhu2025internvl3,cheng2024videollama} have made significant advances, demonstrating remarkable performance in tasks such as dialogue systems, video understanding, and video question answering. Representative models include ChatGPT \cite{hurst2024gpt}, LLaMA \cite{touvron2023llama}, Qwen \cite{bai2023qwen}, and DeepSeek families \cite{liu2024deepseek}. However, existing models primarily focus on surface-level interactions among text, audio, and visual modalities, and still fall short in understanding the fine-grained cross-modal associations within complex audio-visual (A/V) events. Specifically, there is a lack of modeling for non-speech auditory information, such as object collisions, human actions, and environmental background sounds, which are challenging to semantically align with visual content. Moreover, current models exhibit limitations in cross-modal temporal alignment. Although MLLMs can process unimodal inputs, they typically underperform in tasks that require precise synchronization between audio and visual signals, such as A/V alignment detection or temporal localization of audio segments corresponding to specific video frames.

Several multimodal benchmarks \cite{gong2024av,li2024omnibench,benchekroun2023worldsense,zhou2025daily} have been proposed to evaluate the understanding capabilities of MLLMs. However, they exhibit three major limitations: \textbf{(1) Data source bias:} Existing datasets are primarily collected from user-generated content (UGC) platforms such as YouTube, while lacking professionally generated content (PGC) and AI-generated content (AIGC). PGC offers higher-quality A/V content, such as character close-ups, special effects, and refined visual storytelling. AIGC introduces unrealistic A/V content that doesn't exist in the real world \cite{zhangbench,li2023agiqa}. \textbf{(2) Insufficient task coverage:} Current benchmarks mainly focus on fundamental tasks like A/V recognition and reasoning, while paying limited attention to quality-oriented tasks. Some benchmarks \cite{wang2025aigv,wang2025lmm4lmm,zhang2025q} assess MLLMs' visual quality perception but rarely explore A/V quality perception, which evaluates whether MLLMs can perceive the quality or inconsistencies in A/V content, similar to human perception.

In this paper, we introduce \textbf{XGC-AVis}, a novel A/V agent system that improves temporal alignment by interweaving video, audio, subtitles, and descriptions. It identifies relevant time segments, ensuring the MLLM's attention on the most significant content. To comprehensively assess XGC-AVis and MLLMs, we present \textbf{XGC-AVQuiz}, a holistic benchmark designed to evaluate MLLMs in recognition, localization, quality perception, and reasoning when processing real-world and generative A/V content. XGC-AVQuiz incorporates two key innovations: \textbf{(1) Diverse video sources: }XGC-AVQuiz consists of $2232$ A/V samples, including $1102$ PGC videos, $753$ UGC videos, and $377$ AIGC videos. These samples effectively address the data diversity limitations of existing benchmarks through the inclusion of real-world and AIGC scenarios.
\textbf{(2) Multi-level task hierarchy:} The benchmark comprises $2685$ carefully constructed question-answer (QA) pairs, covering $4$ categories and $20$ tasks. The categories are designed with a gradual difficulty progression—from low-level A/V recognition and localization, to high-level A/V reasoning, and comprehensive A/V quality perception. This enables an in-depth assessment of MLLMs' cognitive capabilities across various levels. 

We conduct extensive evaluations on a wide range of MLLMs, including both open-source MLLMs, proprietary MLLMs, and multi-agent systems. XGC-AVis demonstrated the best performance on XGC-AVQuiz and Daily-Omni \cite{zhou2025daily} benchmarks, showcasing its superior ability in handling complex A/V tasks. The results also reveal significant limitations in current MLLMs' ability to understand A/V content. Specifically, most open-source MLLMs struggle with integrating audio, video, and subtitle information, sometimes failing to outperform Vision-Language Models (VLMs). Additionally, MLLMs face challenges in A/V quality perception tasks, indicating considerable potential for further improvement. Furthermore, MLLMs show weaknesses in A/V tasks involving temporal localization.

\vspace{-0.2cm}
\section{Related Works}
\vspace{-0.1cm}
\subsection{Multimodal Large Language Models}
In recent years, MLLMs for audio-visual understanding have made significant progress. Most existing approaches first extract visual embeddings and audio embeddings separately using dedicated visual encoders (e.g., ViT \cite{dosovitskiy2020image}, CLIP-ViT \cite{radford2021learning}), and audio encoders (e.g., BEATs \cite{chen2022beats}, HuBERT \cite{hsu2021hubert}). These embeddings are then concatenated and fed into a large language model for cross-modal reasoning. Some studies \cite{ye2024cat,zhan2024anygpt,wang2025u} further incorporate modality-specific decoders, such as speech or music decoders, to enhance the MLLMs' ability to interpret complex audio content. While this paradigm enables accurate recognition of visual objects, actions, speech, and music, it suffers from a key limitation: poor temporal alignment. In this work, we propose the multi-agent system XGC-AVis, which improves the temporal alignment of audio and video events by interweaving video, audio, subtitles, and audio descriptions.

\subsection{Multimodal Benchmarks}
With the rapid development of MLLMs, a growing number of benchmarks have been proposed. Vision-focused datasets and benchmarks \cite{li2024mvbench,zhang2025q,zhangbench} mainly address perception and understanding tasks involving static images or dynamic videos, while audio-centric datasets and benchmarks \cite{wang2024audiobench,zhu2025does} focus on speech, music, and general sound understanding. However, most benchmarks overlook the importance of joint audio-visual perception. Although several benchmarks have been introduced for audio-visual tasks, they still exhibit key limitations. For instance, AV-Odyssey Bench \cite{gong2024av} and OmniBench \cite{li2024omnibench} focus on static images, while Music-AVQA \cite{li2022learning} and AVQA \cite{yang2022avqa} are domain-specific. Recently, audio-visual benchmarks such as WorldSense \cite{benchekroun2023worldsense} and Daily-Omni \cite{zhou2025daily} have been proposed, providing valuable resources for real-world audio-visual question answering. In contrast, our proposed XGC-AVQuiz incorporates PGC, UGC, and AIGC scenarios, explicitly integrates quality perception tasks, and offers a multi-level task hierarchy that enables deeper insights into MLLMs' limitations and guides their advancement toward comprehensive full-modality understanding and perception.

\begin{figure*}[t]
\begin{center}
\centerline{\includegraphics[width=1\columnwidth]{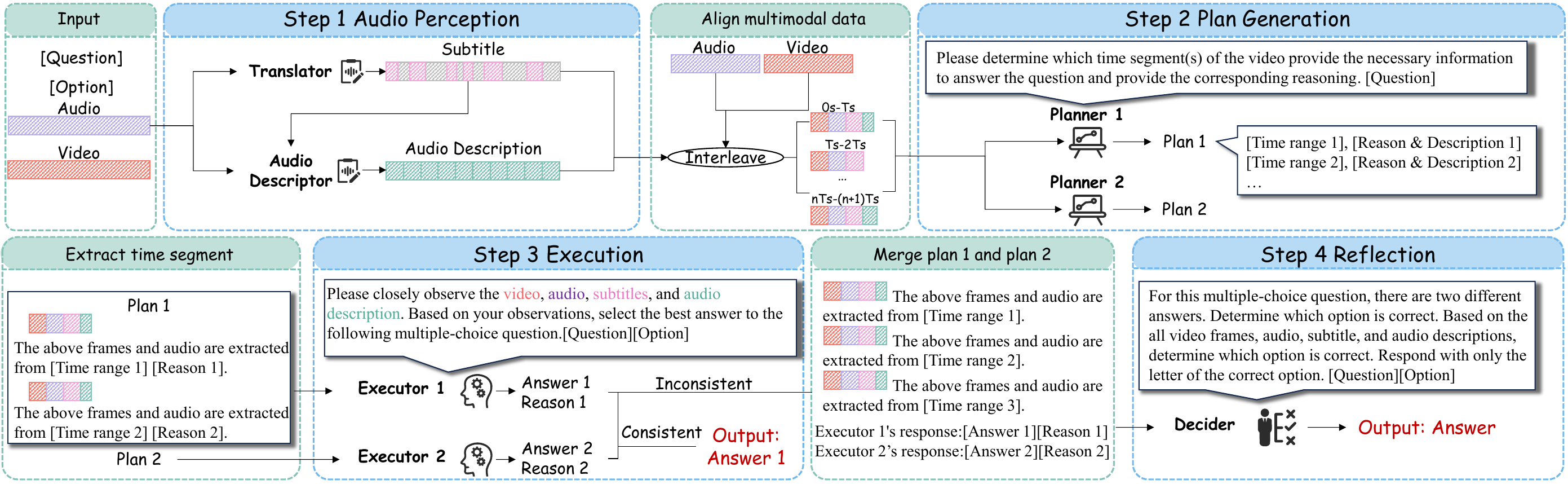}}
\caption{The pipeline of XGC-AVis multi-agent system includes four steps: audio perception, plan generation, execution, and reflection.}
\label{fig:agent}
\end{center}
\vspace{-3em}
\end{figure*}
\vspace{-0.2cm}
\section{XGC-AVis}
\label{sec:XGC-AVis}
\vspace{-0.1cm}
To enhance the perception and understanding capabilities of MLLMs for multimodal information, we design a multi-agent system named XGC-AVis to assist MLLMs in identifying key time points and temporally aligning video and audio events. As shown in Fig. \ref{fig:agent}, this process involves four sequential stages: audio perception, plan generation, execution, and reflection. XGC-AVis first segments the audio into equal-length clips. A translator then converts speech into subtitles, while an audio descriptor detects non-speech events. Video frames, audio segments, subtitles, and audio descriptions are interwoven and temporally aligned to form coherent multimodal units. Next, two planners independently analyze these units to identify key time segments relevant to the question. Each planner outputs the relevant time span and reasoning, forming a targeted answering plan that enhances the efficiency of retrieving important video segments. These plans are passed to dedicated executors, each focusing exclusively on its assigned segment and generating an answer along with a reasoning path using associated subtitles, audio descriptions, and contextual cues. If all executors produce the same answer, it is returned directly. In cases of disagreement, the system merges the conflicting plans, answers, and reasoning and forwards them to a decider to derive the final answer. More details about XGC-AVis can be found in Appendix \ref{sec:XGC-AVis Details}.

In this paper, XGC-AVis employs Deepgram \footnote{\url{https://developers.deepgram.com/}} as the translator, r1-aqa \cite{li2025reinforcement} as the audio descriptor, Aria \cite{li2024aria} and Qwen2.5-Omni \cite{xu2025qwen2} as Planner 1 and Planner 2, and Gemini 2.0 Flash as Executor 1, Executor 2, and the decider.

\vspace{-0.1cm}
\section{XGC-AVQuiz}
\label{sec:XGC-AVQuiz}
\vspace{-0.1cm}
In this section, we first provide a brief overview of XGC-AVQuiz and compare it with existing benchmarks in Section \ref{sec:XGC-AVQuiz1}. Section \ref{sec:XGC-AVQuiz2} describes the video collection and preprocessing pipeline, and Section \ref{sec:XGC-AVQuiz3} details the QA pair annotation process. Specific QA examples of XGC-AVQuiz can be found in Appendix \ref{sec:QA Examples from Each Task}

\subsection{Benchmark Overview and Comparison}
\label{sec:XGC-AVQuiz1}
\begin{figure*}[t]
\begin{center}
\centerline{\includegraphics[width=\columnwidth]{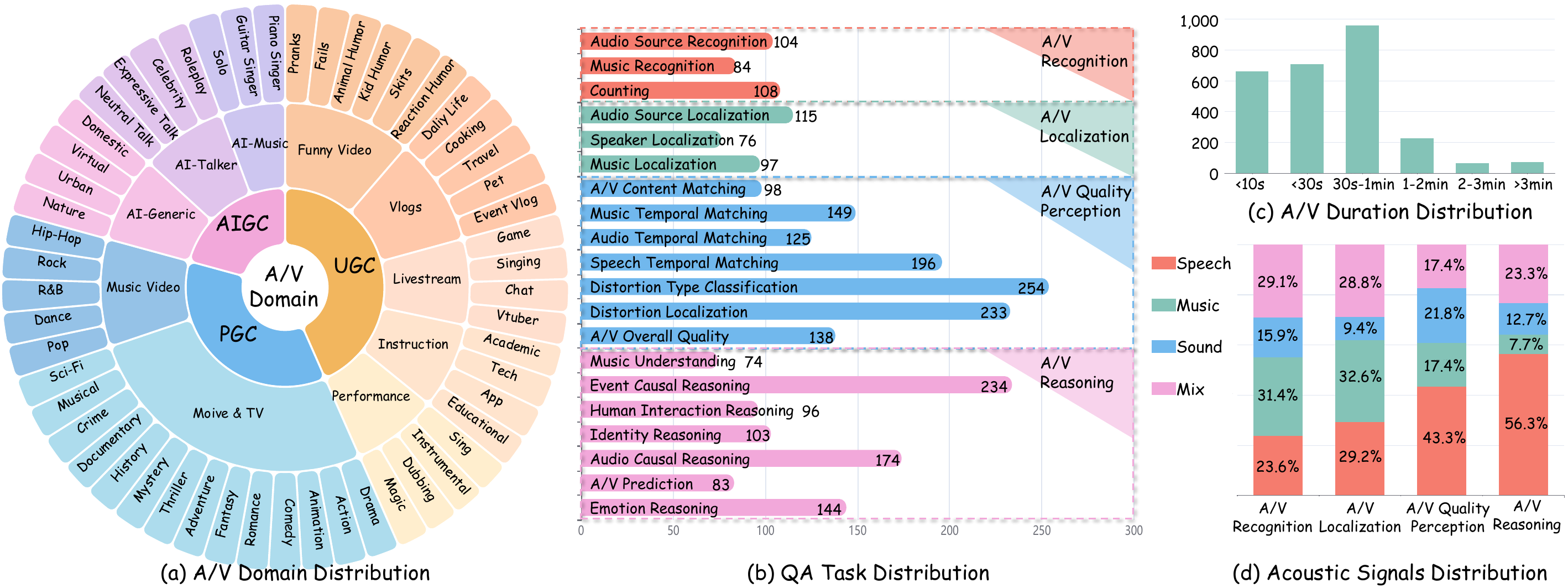}}
\vspace{-0.8em}
\caption{Distribution of videos and question-answer pairs in XGC-AVQuiz.}
\label{fig:overview}
\end{center}
\vspace{-3em}
\end{figure*}

As illustrated in Fig. \ref{fig:overview}, XGC-AVQuiz comprises $2,232$ videos, including $1,102$ PGC videos and $753$ UGC videos representing real-world scenarios, as well as $377$ AIGC videos representing non-realistic scenarios. These videos span $10$ major domains and $53$ categories. The distribution of video lengths is shown in Fig. \ref{fig:overview}(c), where $36\%$ of the videos have durations between 30 seconds and 1 minute, with an average duration of $43s$. 
We design a four-level evaluation framework to comprehensively assess A/V understanding. The first level, A/V recognition, focuses on identifying audio and visual events. Next, A/V localization evaluates the model's ability to pinpoint the temporal or spatial location of A/V events. A/V reasoning requires MLLMs to infer relationships and meanings based on A/V events, demonstrating a deeper level of understanding. Finally, A/V quality perception combines low-level sensory analysis with high-level semantic understanding to assess A/V quality.
Across these four levels, we further define $20$ tasks and collect $2,685$ QA pairs, as shown in Fig.~\ref{fig:overview}(b). 
Based on the type of audio information required by each QA pair, we categorize them into four types: speech, sound, music, and mixed. The distribution of each audio type across different tasks is illustrated in Fig.~\ref{fig:overview}(d).

As summarized in Table~\ref{tab:benchmark}, \textbf{XGC-AVQuiz} is the first benchmark to comprehensively evaluate the multimodal understanding capabilities of MLLMs across both real-world and non-realistic scenarios. Its key contributions include: 
(1) \textbf{Evaluation of AIGC content}: AIGC content can present A/V scenes that are impossible in the real world. XGC-AVQuiz leverages AIGC videos to evaluate whether MLLMs can semantically and temporally align audio and video, and comprehend non-realistic content.
(2) \textbf{Emphasis on A/V Quality Perception}: XGC-AVQuiz evaluates MLLMs' ability to perceive A/V quality through low-quality content. This task requires both low-level sensory abilities (e.g., detecting noise, blur, or latency) and high-level understanding (e.g., assessing semantic consistency and temporal synchronization between audio and video). (3) \textbf{Broader video coverage}: Compared to existing benchmarks with similar numbers of QA pairs, XGC-AVQuiz includes more videos. This provides broader coverage across diverse video types, scenes, and A/V conditions, enabling a more robust assessment of model generalization and robustness.
\begin{table*}[t]
\small
\centering
\belowrulesep=0pt
\aboverulesep=0pt
\renewcommand\arraystretch{1.03}
\renewcommand\tabcolsep{2.5pt}
\vspace{-0.4cm}
\caption{Comparison with existing audio-visual benchmarks. The table compares existing audio-visual benchmarks based on several key features, including \textbf{modality} (A: audio, V: video, I: image), dataset size (\textbf{\#Videos}, \textbf{\#QA Pairs}), \textbf{annotation} method (A: automatic, M: manual), as well as whether the datasets support \textbf{multi-task} QA, include \textbf{non-realistic} content, contain \textbf{quality perception} questions, and offer diverse \textbf{general sound} coverage.}
\resizebox{\linewidth}{!}{
    \minitiny
    \begin{tabular}{c|cccccccc}
    \toprule
    Benchmarks & Modality & \#Video & \#QA Pairs & Annotation & Multi-Task & Non-Realistic Content & Quality Perception & General Sound\\
    \hline
    AVQA \cite{yang2022avqa} & A+V & 57,000 & 57,335 & M & \ding{55} & \ding{55} &\ding{55} & \ding{51}\\
    Music-AVQA \cite{li2022learning} & A+V & 9,288 & 45,867 & M & \ding{55} & \ding{55} &\ding{55} & \ding{55}\\
    OmniBench \cite{li2024omnibench} & A+I & \ding{55} & 1,142 & M & \ding{51} & \ding{55} &\ding{55} & \ding{51} \\
    AV-Odyssey \cite{gong2024av} & A+I & \ding{55} & 4,555 & M & \ding{51} & \ding{55} &\ding{55} & \ding{51} \\
    WorldSense \cite{benchekroun2023worldsense} & A+V & 1,662 & 3,172 & M & \ding{51} & \ding{55} &\ding{55} & \ding{51} \\
    Daily-Omni \cite{zhou2025daily} & A+V & 684 & 1197 & A\&M & \ding{51} & \ding{55} &\ding{55} & \ding{51} \\
    \hline
    \textbf{XGC-AVQuiz (Ours)} & A+V & 2232 & 2685 & M & \ding{51} & \ding{51} &\ding{51} & \ding{51} \\
    \bottomrule
\end{tabular}}
\vspace{-0.2cm}
\label{tab:benchmark}
\end{table*}
\begin{figure*}[t]
\vspace{-0.2cm}
\begin{center}
\centerline{\includegraphics[width=\columnwidth]{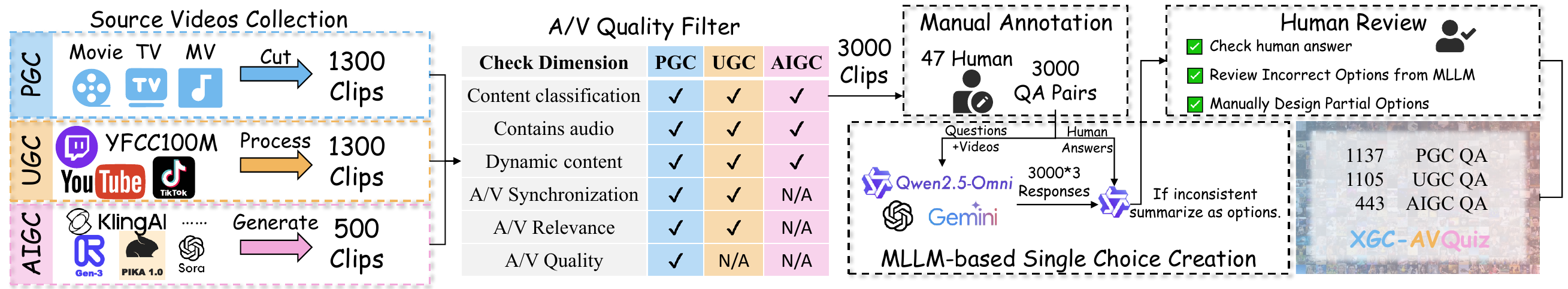}}
\vspace{-0.8em}
\caption{Construction pipeline of the XGC-AVQuiz dataset. (a) Video collection; (b) Quality filtering; (c) Manual annotation; (d) MLLM-based multiple choice creation; (e) Expert review.}
\label{AGAVQA}
\end{center}
\vspace{-3em}
\end{figure*}
\vspace{-0.1cm}
\subsection{Data Collection and Preprocessing}
\label{sec:XGC-AVQuiz2}
\vspace{-0.1cm}
For PGC videos, we first collected $108$ English-language films and TV shows spanning $14$ different themes, and trimmed them into 10-minute clips for annotation of QA pairs. Additionally, we downloaded $100$ music videos in Korean, English, and Japanese to ensure musical diversity. For UGC videos, we gathered $1,000$ short videos from platforms such as YouTube and TikTok, covering game streams, chat sessions, and singing performances. We also collected $100$ livestream recordings from Twitch, including game, chat, and singing. Regarding AIGC videos, we sourced AI-generated content from YouTube featuring virtual avatars speaking and singing. To further enrich audio diversity, we also collected AIGC videos with sound from the AIGC sharing platform Kling \cite{kling}.

To evaluate the A/V quality perception capabilities of MLLMs, we included both high-quality and low-quality A/V content. Low-quality UGC videos were obtained from the UGC A/V quality assessment dataset SJTU-UAV \cite{cao2023subjective}, which contains distortions introduced during user recording. We also simulated livestream-specific glitches, such as audio stuttering, A/V desynchronization, and video freezing, by preprocessing livestream recordings. Low-quality AIGC videos were collected from the AIGC A/V quality assessment dataset AGAVQA-3k \cite{cao2025agav}, which exhibit issues such as A/V misalignment, semantic inconsistency, and unnatural audio.

After collecting the videos, we first categorized all videos and ensured they contained both audio and dynamic visual content. For high-quality samples, we further verified their A/V synchronization, semantic relevance, and overall A/V quality.

\subsection{Annotation Process}
\label{sec:XGC-AVQuiz3}
We recruited $47$ professional annotators to create high-quality QA pairs by considering both audio and video information. Annotators were required to generate QA pairs evenly distributed across all $20$ task types.
All submitted QA pairs were reviewed by domain experts for correctness, clarity, logical consistency, and the necessity of multimodal information. To construct challenging multiple-choice questions, we leveraged MLLMs to generate distractors. Each question and its corresponding video were fed into several MLLMs, and incorrect responses were collected as potential distractors.
Finally, professional annotators refined the distractors based on model-generated errors, and all options were reviewed by experts to ensure quality. This annotation framework, combining expert validation with LLM-assisted distractor generation, guarantees the accuracy of QA pairs while increasing the difficulty of multiple-choice questions, thus posing greater challenges to MLLMs.


\section{Experiment}
In this section, we conduct a comprehensive evaluation of recent MLLMs and our proposed XGC-AVis agent on the XGC-AVQuiz benchmark, and further evaluate XGC-AVis on the Daily-Omni benchmark \cite{zhou2025daily}. In addition, we perform ablation studies on both MLLMs and XGC-AVis to investigate key factors influencing their performance.

\begin{table*}[!t]
\vspace{-0.4cm}
\centering
\belowrulesep=0pt
\aboverulesep=0pt
\renewcommand\arraystretch{1.1}
\renewcommand\tabcolsep{2.5pt}
\caption{Performance on XGC-AVQuiz across four categories. \red{Best} and \blue{second-best} results are highlighted. $\blacktriangle$: VLMs. $\blacklozenge$: open-source OLMs. $\bigstar$: closed-source MLLMs. $\heartsuit$: multi-agent systems.}
\resizebox{\linewidth}{!}{\normalsize 
    \begin{tabular}{l|c|ccc|ccc|ccc|ccc|ccc}
    \toprule
    \multirow{2}{*}{Methods} &\multirow{2}{*}{\shortstack{LLM \\ Size}} & \multicolumn{3}{c|}{A/V Recognition} & \multicolumn{3}{c|}{A/V Localization} & \multicolumn{3}{c|}{A/V Perception} & \multicolumn{3}{c|}{A/V Reasoning} & \multicolumn{3}{c}{Average}\\
    \cline{3-17}
    ~ & ~ & Vid. & +Aud. & +Sub. & Vid. & +Aud. & +Sub. & Vid. & +Aud. & +Sub. & Vid. & +Aud. & +Sub. & Vid. & +Aud. & +Sub. \\
    \hline
    $\blacktriangle$LLaVA-OneVision & 8B & 47.0 & \text{--} & 47.0 & 42.0 & \text{--} & 42.7 & 40.2 & \text{--} & 41.3 & 47.4 & \text{--} & 48.0 & 43.6 & \text{--} & 44.4\\
    $\blacktriangle$InternVL2.5 & 8B & 36.8 & \text{--} & 37.2 & 42.0 & \text{--} & 42.0 & 34.6 & \text{--} & 34.2 & 48.1 & \text{--} & 48.6 & 40.2 & \text{--} & 40.2\\
    $\blacktriangle$InternVL3 & 8B & 48.3 & \text{--} & 47.6 & 50.0 & \text{--} & \blue{50.3} & 40.4 & \text{--} & 40.3 & 52.3 & \text{--} & 52.2 & 46.3 & \text{--} & 46.2\\
    \hdashline
    $\blacklozenge$VideoLLaMA2 & 7B & 42.6 & 48.0 & 53.0 & 41.0 & 41.7 & 39.2 & 42.5 & 40.7 & 43.1 & 39.9 & 40.2 & 51.7 & 41.5 & 41.4 & 46.7\\
    $\blacklozenge$PandaGPT & 7B & 32.4 & 38.2 & 35.8 & 31.3 & 28.8 & 32.3 & 39.6 & 28.9 & 29.4 & 34.9 & 35.4 & 35.5 & 36.4 & 32.1 & 32.5\\
    $\blacklozenge$GroundingGPT & 7B & 38.9 & 40.5 & 40.9 & 35.8 & 37.8 & 35.8 & 30.1 & 27.2 & 29.5 & 37.6 & 36.2 & 48.8 & 34.2 & 32.8 & 38.0\\
    $\blacklozenge$Bubogpt & 7B & 23.0 & 18.2 & 13.5 & 15.6 & 14.9 & 17.7 & 19.4 & 17.4 & 21.2 & 15.1 & 14.2 & 15.3 & 17.9 & 16.2 & 18.0\\
    $\blacklozenge$Unified-IO-2 & 1B & 30.4 & 31.8 & 30.1 & 29.5 & 28.5 & 29.9 & 38.3 & 35.5 & 34.9 & 28.1 & 28.4 & 26.4 & 33.0 & 31.9 & 30.9\\
    $\blacklozenge$Unified-IO-2 XL & 3B & 34.1 & 35.1 & 35.1 & 24.7 & 25.7 & 25.3 & 28.0 & 28.9 & 27.5 & 27.4 & 29.8 & 27.4 & 28.1 & 29.6 & 28.1\\
    $\blacklozenge$Unified-IO-2 XXL & 8B & 32.4 & 36.5 & 33.4 & 27.1 & 27.4 & 29.5 & 38.7 & 34.2 & 37.3 & 30.4 & 33.0 & 31.1 & 34.0 & 33.3 & 33.9\\
    $\blacklozenge$Video-SALMONN & 7B & 32.4 & 34.5 & 33.8 & 33.7 & 36.5 & 35.4 & 24.7 & 26.2 & 29.2 & 35.5 & 36.0 & 44.7 & 30.2 & 31.5 & 35.6\\
    $\blacklozenge$Qwen2.5-Omni & 7B & 47.6 & 55.4 & 53.7 & 40.6 & 37.5 & 33.0 & 37.1 & \blue{45.1} & 41.8 & 45.0 & 57.9 & 55.1 & 41.3 & 49.8 & 46.7\\
    \hdashline
    $\bigstar$ChatGPT-4o & \text{--} & 44.6 & \text{--} & 44.9 & 47.2 & \text{--} & 47.9 & 23.4 & \text{--} & 36.6 & 50.4 & \text{--} & 60.1 & 37.4 & \text{--} & 46.7\\
    $\bigstar$Claude 3.7 Sonnet & \text{--} & 31.8 & \text{--} & 28.7 & 33.7 & \text{--} & 29.5 & 15.4 & \text{--} & 15.1 & 31.9 & \text{--} & 49.6 & 24.8 & \text{--} & 29.8\\
    $\bigstar$Gemini 2.0 Flash & \text{--} & 45.9 & 54.4 & \blue{55.7} & 43.4 & 49.3 & 49.3 & 34.3 & 40.2 & 38.1 & 48.5 & 65.9 & \blue{68.1} & 41.3 & \blue{51.4} & 51.3\\
    \hdashline
    $\heartsuit$Daliy-Omni & \text{--} & & 38.9 & & & 25.3 & & & 26.1 & & & 43.7 & & & 33.4\\
    \rowcolor{mygray} $\heartsuit$\textbf{XGC-AVis (Ours)} & \text{--} & & \red{59.5} & & & \red{51.7} & & & \red{51.0} & & & \red{69.3} & & & \red{58.2} &  \\
    \rowcolor{mygray} \textit{Improvement} & \text{--} & & \textit{+3.8\%} & & & \textit{+1.4\%} & & & \textit{+5.9\%} & & & \textit{+1.2\%} & & & \textit{+6.8\%} &  \\
    \bottomrule
\end{tabular}}
\label{tab:mainresult}
\end{table*}
\begin{table*}[!t]
\vspace{-0.6cm}
\centering
\belowrulesep=0pt
\aboverulesep=0pt
\renewcommand\arraystretch{1.1}
\renewcommand\tabcolsep{2.5pt}
\caption{Performance on Daily-omni across different different question domains and video durations. \red{Best} and \blue{second-best} results are highlighted. $\blacklozenge$: open-source OLMs. $\bigstar$: closed-source MLLMs. $\heartsuit$: multi-agent systems.}
\resizebox{\linewidth}{!}{\normalsize
    \begin{tabular}{l|ccc ccc| cc | c}
    \toprule
    \multirow{2}{*}{Methods} &\multirow{2}{*}{\shortstack{A/V Event \\ Alignment}}  & \multirow{2}{*}{\shortstack{Comparative}} & \multirow{2}{*}{\shortstack{Context\\Understanding}} & \multirow{2}{*}{\shortstack{Event\\Sequence}} & \multirow{2}{*}{\shortstack{Inference}} & \multirow{2}{*}{\shortstack{Reasoning}} & \multirow{2}{*}{\shortstack{\SI{30}{\second}\\Subset}} & \multirow{2}{*}{\shortstack{$60s$\\Subset}} & \multirow{2}{*}{\shortstack{Avg}} \\
    ~ & ~ & ~ & ~ & ~ & ~ & ~ & ~ & ~ & ~ \\ 
    \hline
    $\blacklozenge$Unified-IO-2XXL (8B) & 44.1 & 51.2 & 38.9 & 40.5 & 57.8 & 61.7 & 46.7 & 48.4 & 47.5\\
    $\blacklozenge$VideoLLaMA2 (7B) &35.7 & 35.9 & 35.8 & 31.7 & 40.9 & 34.3 & 38.0 & 31.8  & 35.2\\
    $\blacklozenge$Qwen2.5-Omni (7B) & 25.6 & 31.3 & 26.4 & 25.8 & 35.1 & 29.7 & 26.7 & 30.0 & 28.2\\
    $\bigstar$ChatGPT-4o  & 47.9 & 62.6 & 52.3 & 52.6 & 66.2 & 66.3 & 55.6 & 57.5 & 56.5 \\
    $\bigstar$Gemini 2.0 Flash Lite  & 55.0 & 64.9 & 58.0 & 54.3 & 74.0 & 72.0 & 62.4 & 60.0  & 61.3\\
    $\bigstar$Gemini 2.0 Flash & \blue{62.2}  & \blue{73.3}  & \blue{63.7}  & \blue{63.7}  & 76.6  & \blue{75.4}  & \blue{67.2}  & \blue{68.6}  & \blue{67.8} \\
    \hdashline
    $\heartsuit$Daily-Omni & 51.7 & 68.7  & 60.1  & 59.3 & \blue{78.6}  & 71.4  & 64.0 & 59.3  & 61.8 \\
    \rowcolor{mygray}$\heartsuit$\textbf{XGC-AVis (Ours)} & \red{63.5}  & \red{77.1}  & \red{68.4}  & \red{64.4}  & \red{85.1}  & \red{82.3}  & \red{71.6}  & \red{71.5}  & \red{71.5} \\    
    \rowcolor{mygray} \textit{Improvement} & \textit{+1.3\%} & \textit{+3.8\%} & \textit{+4.7\%} & \textit{+0.7\%} & \textit{+6.5\%} & \textit{+6.9\%} & \textit{+4.4\%} & \textit{+2.9\%} & \textit{+3.7\%} \\
    \bottomrule
\end{tabular}}
\label{tab:daily}
\end{table*}

\subsection{Experimental Settings}
\vspace{-0.1cm}
Our evaluation covers four types of MLLMs: (1) Open-source VLMs: LLaVA-OneVision \cite{li2024llava}, InternVL2.5 \cite{chen2024expanding}, and InternVL3 \cite{zhu2025internvl3}. (2) Open-source omni-language models (OLMs): VideoLLaMA2 \cite{cheng2024videollama}, PandaGPT \cite{su2023pandagpt}, GroundingGPT \cite{li2024groundinggpt}, Bubogpt \cite{zhao2023bubogpt}, Unified-IO-2 \cite{lu2024unified}, Video-SALMONN \cite{sun2024video}, Qwen2.5-Omni  \cite{xu2025qwen2}. (3) Closed-source MLLMs: GPT-4o \cite{hurst2024gpt}, Claude 3.7 Sonnet \cite{claude3}, and Gemini 2.0 Flash \cite{gemini20flash}. (4) Multi-agent systems: Daliy-Omni \cite{zhou2025daily} and our XGC-AVis. We further consider three input settings: video only, video with audio, and video with audio and subtitles. More experiment details are in Appendix \ref{sec:MLLM Experiment Details} and \ref{sec:Details on LLM-assisted Evaluation}.


\begin{figure*}[!t]
\begin{center}
\vspace{-0.4cm}
\centerline{\includegraphics[width=1\columnwidth]{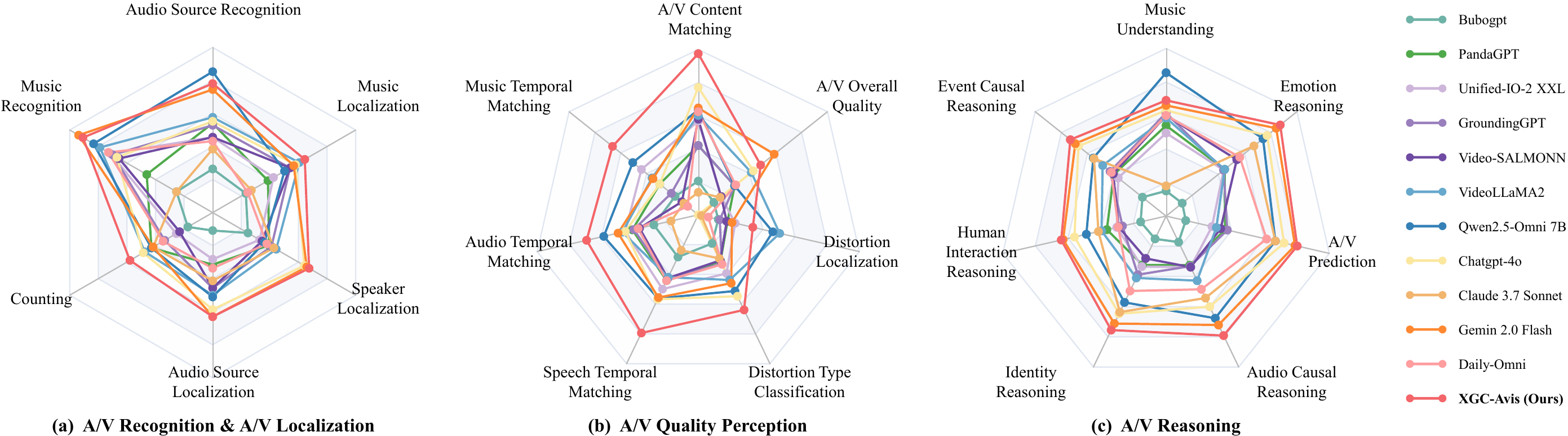}}
\caption{MLLMs' accuracy over different A/V task categories.}
\vspace{-1cm}
\label{fig:overall_leida}
\end{center}
\end{figure*}
\begin{table*}[!t]
\vspace{-0.4cm}
\centering
\belowrulesep=0pt
\aboverulesep=0pt
\renewcommand\arraystretch{1.1}
\renewcommand\tabcolsep{2.5pt}
\caption{Performance on XGC-AVQuiz across different video durations. \red{Best} and \blue{second-best} results are highlighted. $\blacklozenge$: open-source OLMs. $\bigstar$: closed-source MLLMs. $\heartsuit$: multi-agent systems.}
\resizebox{\linewidth}{!}{\normalsize
    \begin{tabular}{l|c|ccc|ccc|ccc|ccc|ccc}
    \toprule
    \multirow{2}{*}{Methods} & \multirow{2}{*}{\shortstack{LLM \\ Size}}  & \multicolumn{3}{c|}{$<10\mathrm{s}$} & \multicolumn{3}{c|}{$10\mathrm{s}-30\mathrm{s}$} & \multicolumn{3}{c|}{$30\mathrm{s}-1\mathrm{min}$} & \multicolumn{3}{c|}{$1\mathrm{min}-2\mathrm{min}$} & \multicolumn{3}{c}{$>2\mathrm{min}$}\\
    \cline{3-17}
    ~ & ~ & Vid. & +Aud. & +Sub. & Vid. & +Aud. & +Sub. & Vid. & +Aud. & +Sub. & Vid. & +Aud. & +Sub. & Vid. & +Aud. & +Sub. \\
    \hline
    $\blacktriangle$LLaVA-OneVision & 8B & 37.8 & \text{--} & 37.2 & 47.3 & \text{--} & 48.6 & 43.2 & \text{--} & 44.6 & 47.3 & \text{--} & 48.7 & 49.2 & \text{--} & 48.5\\
    $\blacktriangle$InternVL2.5 & 8B & 38.4 & \text{--} & 40.5 & 40.5 & \text{--} & 39.8 & 37.9 & \text{--} & 36.5 & 50.0 & \text{--} & 50.4 & 48.5 & \text{--} & 50.8\\											
    $\blacktriangle$InternVL3 & 8B & 43.4 & \text{--} & 42.9 & 49.7 & \text{--} & 50.3 & 44.8 & \text{--} & 44.0 & 52.7 & \text{--} & 53.1 & 43.2 & \text{--} & 45.5\\
    \hdashline
    $\blacklozenge$VideoLLaMA2 & 7B & 40.5 & 39.0 & 41.7 & 41.4 & 42.7 & 48.0 & 41.7 & 42.5 & 47.3 & 42.4 & 40.2 & 49.6 & 43.2 & 40.9 & 54.5\\
    $\blacklozenge$PandaGPT & 7B & 30.4 & 30.1 & 30.8 & 38.4 & 39.1 & 38.0 & 39.7 & 27.0 & 28.1 & 31.7 & 35.3 & 38.4 & 38.6 & 36.4 & 33.3\\
    $\blacklozenge$GroundingGPT & 7B & 32.5 & 32.9 & 34.6 & 40.3 & 39.1 & 46.2 & 30.0 & 27.6 & 32.1 & 40.2 & 37.1 & 48.2 & 28.8 & 29.5 & 35.6\\
    $\blacklozenge$Bubogpt & 7B & 15.3 & 14.2 & 15.4 & 21.2 & 18.5 & 16.7 & 17.4 & 16.8 & 21.4 & 18.8 & 12.1 & 14.3 & 15.9 & 15.9 & 19.7\\
    $\blacklozenge$Unified-IO-2 & 1B &	30.7 & 31.0 & 31.3 & 30.5 & 31.5 & 30.5 & 38.4 & 34.8 & 32.4 & 29.5 & 28.6 & 29.9 & 25.8 & 23.5 & 22.7\\
    $\blacklozenge$Unified-IO-2 XL & 3B & 30.5 & 31.0 & 30.2 & 33.9 & 35.0 & 34.3 & 22.9 & 24.1 & 22.7 & 26.3 & 33.9 & 27.7 & 25.8 & 25.8 & 23.5\\
    $\blacklozenge$Unified-IO-2 XXL & 8B & 38.2 & 37.2 & 39.4 & 31.8 & 33.5 & 32.3 & 34.9 & 31.2 & 33.2 & 28.1 & 33.0 & 28.6 & 27.3 & 29.5 & 29.5\\
    $\blacklozenge$Video-SALMONN & 7B & 33.7 & 34.3 & 33.2 & 34.5 & 36.4 & 41.2 & 22.4 & 24.4 & 31.4 & 34.4 & 36.6 & 50.0 & 38.6 & 34.8 & 23.5\\
    $\blacklozenge$Qwen2.5-Omni & 7B & 42.9 & 39.1 & 39.7 & 45.5 & 55.1 & 51.7 & 34.9 & \blue{51.8} & 46.4 & 49.6 & 52.7 & 50.0 & 43.9 & 54.5 & 50.8\\
    \hdashline
    $\bigstar$ChatGPT-4o & \text{--} & 43.7 & \text{--} & 49.8 & 42.9 & \text{--} & 47.6 & 23.1 & \text{--} & 37.7 & 53.1 & \text{--} & \blue{63.8} & 53.8 & \text{--} & \blue{62.1}\\
    $\bigstar$Claude 3.7 Sonnet & \text{--} & 30.5 & \text{--} & 16.6 & 27.8 & \text{--} & 31.2 & 14.7 & \text{--} & 29.1 & 33.9 & \text{--} & 53.1 & 37.1 & \text{--} & 53.8\\
    $\bigstar$Gemini 2.0 Flash & \text{--} & 48.5 & \blue{56.3} & 56.0 & 45.8 & 57.3 & \blue{58.1} & 30.7 & 40.5 & 39.4 & 50.4 & 62.1 & 62.9 & 43.9 & 56.8 & 59.1\\
    \hdashline
    $\heartsuit$Daliy-Omni & \text{--} & & 34.7 & & & 36.6 & & & 26.6 & & & 42.0 & & & 43.9\\
    \rowcolor{mygray} $\heartsuit$\textbf{XGC-AVis (Ours)} & \text{--} & & \red{57.1} & & & \red{62.1} & & & \red{53.1} & & & \red{67.4} & & & \red{63.6} &  \\
    \rowcolor{mygray} \textit{Improvement} & \text{--} & & \textit{+0.8\%} & & & \textit{+4.0\%} & & & \textit{+1.3\%} & & & \textit{+3.6\%} & & & \textit{+1.5\%} &  \\
    \bottomrule
\end{tabular}}
\vspace{-1em}
\label{tab:videoduration}
\end{table*}
\begin{table*}[!t]
\vspace{-0.3cm}
\centering
\belowrulesep=0pt
\aboverulesep=0pt
\renewcommand\arraystretch{1.13}
\renewcommand\tabcolsep{2.5pt}
\caption{Performance on XGC-AVQuiz across different audio types. \red{Best} and \blue{second-best} results are highlighted. $\blacklozenge$: open-source OLMs. $\bigstar$: closed-source MLLMs. $\heartsuit$: multi-agent systems.}
\resizebox{\linewidth}{!}{
    \begin{tabular}{l|c|ccc|ccc|ccc|ccc}
    \toprule
    \multirow{2}{*}{Methods} &\multirow{2}{*}{\shortstack{LLM \\ Size}}  & \multicolumn{3}{c|}{Speech} & \multicolumn{3}{c|}{Sound} & \multicolumn{3}{c|}{Music} & \multicolumn{3}{c}{Mix} \\
    \cline{3-14}
    ~ & ~ & Video & +Audio & +Subtitle & Video & +Audio & +Subtitle & Video & +Audio & +Subtitle & Video & +Audio & +Subtitle \\
    \hline
    $\blacktriangle$LLaVA-OneVision & 8B & 40.7 & \text{--} & 39.8 & 38.5 & \text{--} & 43.9 & 40.0 & \text{--} & 37.4 & 40.7 & \text{--} & 40.4\\
    $\blacktriangle$InternVL2.5 & 8B & 45.3 & \text{--} & 44.8 & 46.8 & \text{--} & 46.3 & 48.0 & \text{--} & \blue{49.0} & 46.9 & \text{--} & 46.9\\
    $\blacktriangle$InternVL3 & 8B & 42.8 & \text{--} & 44.5 & 40.3 & \text{--} & 39.9 & 47.3 & \text{--} & 47.3 & 44.7 & \text{--} & 45.2\\
    \hdashline
    $\blacklozenge$VideoLLaMA2 & 7B & 41.5 & 39.6 & 48.6 & 41.4 & 44.1 & 46.1 & 41.5 & 43.2 & 43.9 & 41.3 & 41.6 & 45.5\\
    $\blacklozenge$PandaGPT & 7B & 38.8 & 31.5 & 32.0 & 34.3 & 34.7 & 33.4 & 33.1 & 32.5 & 32.5 & 35.5 & 31.1 & 32.8\\
    $\blacklozenge$GroundingGPT & 7B & 32.1 & 30.9 & 39.5 & 35.6 & 37.2 & 39.4 & 34.8 & 33.8 & 36.1 & 36.5 & 32.8 & 35.1\\
    $\blacklozenge$Bubogpt & 7B & 18.4 & 17.6 & 18.9 & 12.9 & 15.1 & 12.2 & 24.1 & 15.5 & 19.8 & 16.0 & 14.6 & 19.2\\
    $\blacklozenge$Unified-IO-2 & 1B & 34.6 & 33.8 & 31.6 & 28.1 & 27.4 & 25.4 & 34.2 & 33.8 & 35.1 & 32.8 & 30.1 & 30.7\\
    $\blacklozenge$Unified-IO-2 XL & 3B & 26.2 & 27.7 & 26.7 & 27.6 & 30.3 & 26.9 & 35.9 & 34.8 & 35.7 & 26.1 & 28.5 & 25.6\\
    $\blacklozenge$Unified-IO-2 XXL & 8B & 34.0 & 32.5 & 33.3 & 31.6 & 32.1 & 31.6 & 38.7 & 36.6 & 40.9 & 31.9 & 33.4 & 31.4\\
    $\blacklozenge$Video-SALMONN & 7B & 29.7 & 30.6 & 38.8 & 31.6 & 35.9 & 35.0 & 28.6 & 28.6 & 30.5 & 31.2 & 32.4 & 33.6\\
    $\blacklozenge$Qwen2.5-Omni & 7B & 37.8 & 51.9 & 48.5 & 48.8 & 46.3 & 47.0 & 46.5 & 47.5 & 41.5 & 38.7 & \blue{49.9} & 46.9\\
    \hdashline
    $\bigstar$ChatGPT-4o & \text{--} & 32.4 & \text{--} & 49.7 & 49.4 & \text{--} & 51.4 & 40.9 & \text{--} & 43.2 & 35.7 & \text{--} & 39.9\\
    $\bigstar$Claude 3.7 Sonnet & \text{--} & 20.4 & \text{--} & 42.0 & 25.4 & \text{--} & 16.9 & 31.6 & \text{--} & 19.4 & 27.7 & \text{--} & 23.3\\
    $\bigstar$Gemini 2.0 Flash & \text{--} & 38.7 & \blue{54.1} & 53.3 & 47.7 & 55.7 & \blue{56.1} & 47.3 & 45.4 & 45.2 & 37.2 & 47.7 & 48.7\\
    \hdashline
    $\heartsuit$Daliy-Omni & \text{--} & & 36.7 & & & 38.1 & & & 27.5 & & & 27.7\\
    \rowcolor{mygray} $\heartsuit$\textbf{XGC-AVis (Ours)} & \text{--} & & \red{62.7} & & & \red{57.5} & & & \red{51.6} & & & \red{54.8} & \\
    \rowcolor{mygray} \textit{Improvement} & \text{--} & & \textit{+8.6\%} & & & \textit{+1.4\%} & & & \textit{+2.6\%} & & & \textit{+4.9\%} & \\
    \bottomrule
\end{tabular}}
\vspace{-1.5em}
\label{tab:audiotypes}
\end{table*}

\subsection{Main Result}
Table \ref{tab:mainresult} compares the performance of various MLLMs and multi-agent systems on the four audio-visual tasks in the XGC-AVQuiz benchmark. Our findings offer several key insights into the current state of MLLMs in A/V understanding. \textbf{First, open-source VLMs rely solely on visual and textual inputs, which limits their ability to perform multimodal tasks effectively.} In Table \ref{tab:mainresult}, we utilize the Deepgram API to transcribe audio into subtitles and feed them to VLMs alongside video. However, this does not significantly improve accuracy compared with video-only input, and VLMs underperform Qwen2.5-Omni in most audio-visual tasks.
These results highlight VLMs' inability to leverage subtitle cues and emphasize the necessity of incorporating audio for more comprehensive and accurate A/V understanding.


\textbf{Second, most open-source OLMs struggle to integrate audio, video, and subtitle information and fail to surpass VLMs.} Adding audio or subtitles negatively impacts the average accuracy of most open-source OLMs. Only Qwen2.5-Omni and VideoLLaMA2 effectively leverage audio or subtitle information to achieve higher average accuracy than open-source VLMs. Overall, these results indicate that many open-source OLMs remain heavily dependent on visual inputs, and simple concatenation of audio or subtitle signals is insufficient for effective audio-visual integration.

\textbf{Third, closed-source MLLMs, such as Gemini 2.0 Flash, show poor performance on quality perception tasks compared to open-source OLMs like Qwen2.5-Omni.} For ChatGPT-4o and Claude 3.7 Sonnet, we follow the official guidelines and use ChatGPT-Audio and Deepgram to generate audio descriptions or subtitles as auxiliary inputs. Results show that for A/V recognition and localization tasks, these inputs offer limited cues. Although Gemini 2.0 Flash achieves the highest average performance among both closed-source and open-source OLMs, its performance on the A/V quality perception task still lags behind the open-source OLM Qwen2.5-Omni, indicating that A/V quality perception remains a key challenge with room for further improvement.

Finally, as shown in Tables \ref{tab:mainresult} and \ref{tab:daily}, \textbf{our proposed XGC-AVis achieves state-of-the-art results, outperforming Gemini 2.0 Flash by 6.8\% on XGC-AVQuiz and 3.7\% on Daily-Omni in average accuracy.} Moreover, it surpasses the Daily-Omni agent, demonstrating the stronger robustness and transferability of XGC-AVis.

Fig.~\ref{fig:overall_leida} illustrates MLLMs' accuracy across different A/V task categories. XGC-AVis, which employs Gemini 2.0 Flash as the decider, achieves higher accuracy than Gemini 2.0 Flash on all $20$ A/V tasks except the overall A/V quality task. In the A/V recognition and localization categories, most models perform poorly on audio counting. This indicates that although they are capable of capturing semantic information, they struggle to temporally localize A/V events. In the quality perception category, their weakness in distortion localization further suggests difficulty in modeling temporal variations in A/V quality, posing challenges for quality assessment and distortion detection. \textbf{These results reveal persistent challenges in temporal localization and variation modeling.} Finally, in the reasoning category, most MLLMs show consistent performance, while Qwen2.5-Omni surpasses Gemini 2.0 Flash in music understanding, indicating a stronger ability to perceive emotional cues. 



\begin{table}[!t]
\vspace{-0.4cm}
\centering
\belowrulesep=0pt
\aboverulesep=0pt
\renewcommand\arraystretch{1.13}
\renewcommand\tabcolsep{6pt}
\caption{Performance on XGC-AVQuiz across difference video content. \red{Best} and \blue{second-best} results are highlighted. $\blacklozenge$: open-source OLMs. $\bigstar$: closed-source MLLMs. $\heartsuit$: multi-agent systems.}
\resizebox{\linewidth}{!}{\small
    \begin{tabular}{l|c|ccc|ccc|ccc}
    \toprule
    \multirow{2}{*}{Methods} &\multirow{2}{*}{\shortstack{LLM \\ Size}}  & \multicolumn{3}{c|}{PGC} & \multicolumn{3}{c|}{UGC} & \multicolumn{3}{c}{AIGC} \\
    \cline{3-11}
    ~ & ~ & Video & +Audio & +Subtitle & Video & +Audio & +Subtitle & Video & +Audio & +Subtitle\\
    \hline
    $\blacktriangle$LLaVA-OneVision & 8B & 43.4 & \text{--} & 44.0 & 44.1 & \text{--} & 45.3 & 42.7 & \text{--} & {42.9}\\
    $\blacktriangle$InternVL2.5 & 8B & 42.9 & \text{--} & 43.7 & 39.2 & \text{--} & 37.7 & 35.9 & \text{--} & 37.5\\
    $\blacktriangle$InternVL3 & 8B & 49.0 & \text{--} & 48.5 & 45.4 & \text{--} & 45.4 & 41.8 & \text{--} & 42.4\\
    \hdashline
    $\blacklozenge$VideoLLaMA2 & 7B & 37.5 & 38.6 & 48.3 & 44.6 & 44.4 & \blue{46.6} & 43.8 & 41.1 & 42.7\\
    $\blacklozenge$PandaGPT & 7B & 32.8 & 33.2 & 34.6 & 39.9 & 29.2 & 28.9 & 36.6 & 36.6 & 36.1\\
    $\blacklozenge$GroundingGPT & 7B & 35.5 & 34.4 & 43.6 & 32.0 & 31.0 & 32.8 & 35.7 & 33.4 & 36.3\\
    $\blacklozenge$Bubogpt & 7B & 15.7 & 14.3 & 14.2 & 18.7 & 17.4 & 21.3 & 21.4 & 17.8 & 19.4\\
    $\blacklozenge$Unified-IO-2 & 1B & 27.2 & 26.6 & 25.5 & 38.6 & 36.5 & 34.8 & 34.3 & 34.3 & 35.2\\
    $\blacklozenge$Unified-IO-2 XL & 3B & 25.9 & 29.3 & 25.9 & 25.3 & 25.2 & 26.2 & 40.6 & 41.1 & 38.4\\
    $\blacklozenge$Unified-IO-2 XXL & 8B & 27.9 & 30.3 & 28.0 & 36.7 & 33.3 & 36.1 & 42.7 & 41.3 & 43.8\\
    $\blacklozenge$Video-SALMONN & 7B & 33.3 & 34.1 & 41.1 & 26.1 & 27.8 & 30.7 & 32.3 & 34.3 & 33.9\\
    $\blacklozenge$Qwen2.5-Omni & 7B & 43.4 & 55.8 & 54.2 & 35.4 & 44.3 & 38.1 & \blue{51.0} & 48.1 & 48.8\\
    \hdashline
    $\bigstar$ChatGPT-4o & \text{--} & 46.6 & \text{--} & 55.0 & 27.5 & \text{--} & 37.0 & 38.6 & \text{--} & 49.7\\
    $\bigstar$Claude 3.7 Sonnet & \text{--} & 28.9 & \text{--} & 43.6 & 21.4 & \text{--} & 24.1 & 22.3 & \text{--} & 8.6\\
    $\bigstar$Gemini 2.0 Flash & \text{--} & 44.4 & 59.3 & \blue{61.6} & 35.7 & 43.7 & 41.7 & 47.6 & 50.6 & 49.2\\
    \hdashline
    $\heartsuit$Daliy-Omni & \text{--} & & 41.7 & & & 22.9 & & & 38.1 & \\
    \rowcolor{mygray} $\heartsuit$\textbf{XGC-AVis (Ours)} & \text{--} & & \red{63.8} & & & \red{54.7} & & & \red{52.6} & \\
    \rowcolor{mygray} \textit{Improvement} & \text{--} & & \textit{+2.2\%} & & & \textit{+8.1\%} & & & \textit{+1.6\%} & \\
    \bottomrule
\end{tabular}}
\vspace{-0.5cm}
\label{tab:videocontent}
\end{table}
\subsection{Ablation Study of XGC-AVQuiz}
\textbf{Impact of Video Duration.}
Table \ref{tab:videoduration} illustrates the impact of video durations on MLLMs' accuracy under three input configurations: video-only, video with audio, and video with audio and subtitles. The following insights can be drawn from the results: (1) Our proposed XGC-AVis consistently outperforms other MLLMs, with its largest gain of 4.0\% over Gemini 2.0 Flash on $10$–$30$ second videos. (2) As video duration increases, ChatGPT-4o gradually surpasses Gemini 2.0 Flash, showing showing a stronger ability for long content. However, XGC-AVis, built upon Gemini 2.0 Flash, further improves accuracy on videos longer than $30$ seconds by leveraging its planner to accurately locate question-relevant segments, thereby enhancing temporal localization. This enables it to surpass both Qwen2.5-Omni and ChatGPT-4o.

\textbf{Impact of Audio Type.}
Table \ref{tab:audiotypes} compares the impact of four audio types (speech, sound, music, and mix) on MLLMs' performance. XGC-AVis achieves the best performance across all audio types, with notable improvements of 8.6\% and 4.9\% over Gemini 2.0 Flash on speech and mix, respectively. Although XGC-AVis leverages audio descriptors to aid non-speech sound understanding, significant challenges remain in handling sound and music. 

When evaluating both open-source and closed-source MLLMs, we observe: (1) Even for speech, the addition of subtitles may reduce performance, reflecting the challenge MLLMs face in aligning visual content with subtitles. (2) Closed-source MLLMs outperform open-source MLLMs on speech and audio, but lag behind on music and mix. Notably, Gemini 2.0 Flash performs best with video-only input on music, underscoring its limitations in handling music. (3) ChatGPT-4o and Claude 3.7 Sonnet rely on ChatGPT-Audio and Deepgram to acquire audio descriptions and subtitles, respectively. For Claude 3.7 Sonnet, subtitles improve performance only on speech, while ChatGPT-4o benefits from audio descriptions across all audio types. This demonstrates that audio descriptions convey richer acoustic cues, such as tone, emotion, and environmental sounds, that are crucial for multimodal understanding. Unlike subtitles, they capture information beyond textual content, underscoring the importance of comprehensive acoustic signals for effective multimodal comprehension.

\textbf{Impact of Video Content.} As shown in Table \ref{tab:videocontent}, XGC-AVis achieves the best performance across all three video content types. The largest gain is observed on UGC, where it surpasses the second-best model by 8.1\%. This is partly because we introduced low-quality videos into UGC and AIGC to test MLLMs' quality perception. Since VideoLLaMA2 and Qwen2.5-Omni outperform Gemini 2.0 Flash in quality perception, Gemini 2.0 Flash does not achieve the best results on UGC and AIGC videos. By integrating the insights of Planner 1 and Planner 2, XGC-AVis enhances Gemini 2.0 Flash's ability to perceive quality, thereby improving its overall performance.
\begin{table*}[!t]
\vspace{-0.4cm}
\centering
\belowrulesep=0pt
\aboverulesep=0pt
\renewcommand\arraystretch{1.1}
\renewcommand\tabcolsep{2pt}
\caption{Ablation studys of XGC-AVis on XGC-AVQuiz. \red{Best} results are highlighted.}
\resizebox{\linewidth}{!}{\normalsize 
    \begin{tabular}{l|cccc|cccc|cccc}
    \toprule
    \multirow{3}{*}{Methods} & \multicolumn{4}{c|}{Question Type} & \multicolumn{4}{c|}{Video Duration} & \multicolumn{4}{c}{Audio Type} \\
    \cline{2-13}
    ~ & \multirow{2}{*}{\shortstack{A/V \\ Recognition}} & \multirow{2}{*}{\shortstack{A/V\\Localization}} & \multirow{2}{*}{\shortstack{A/V\\Perception}} & \multirow{2}{*}{\shortstack{A/V\\Reasoning}} & \multirow{2}{*}{\shortstack{$30\mathrm{s}$\\Subset}} & \multirow{2}{*}{\shortstack{$1\mathrm{min}$\\Subset}} & \multirow{2}{*}{\shortstack{$2\mathrm{min}$\\Subset}} & \multirow{2}{*}{\shortstack{$5\mathrm{min}$\\Subset}} & \multirow{2}{*}{speech} & \multirow{2}{*}{audio} & \multirow{2}{*}{music} & \multirow{2}{*}{mix} \\
    ~ & ~ &  ~ &  ~ &  ~ &  ~ &  ~ &  ~ &  ~ &  ~ &  ~ &  ~ &\\
    \hline
    Gemini 2.0 Flash  & 54.4 & 49.3 & 40.2 & 65.9 & 56.9 & 40.5 & 62.1 & 56.8 & 54.1 & 55.7 & 45.4 & 47.7\\
    XGC-AVis w/o Planner 2 & 57.1 & 48.3 & 43.5 & 65.7 & 53.1 & 49.8 & 61.6 & 60.6 & 56.9 & 51.9 & 49.2 & 49.1\\
    XGC-AVis w/o Planner 1  & 48.0 & 39.9 & 41.4 & 52.5 & 44.9 & 45.5 & 48.7 & 51.5 & 49.3 & 45.9 & 42.4 & 41.1\\
    \rowcolor{mygray} $\heartsuit$\textbf{XGC-AVis (Ours)} & \red{59.5} & \red{51.7} & \red{51.0} & \red{69.3} & \red{59.7} & \red{53.1} & \red{67.4} & \red{63.6} & \red{62.7} & \red{57.5} & \red{51.6} & \red{54.8}\\
    \bottomrule
\end{tabular}}
\vskip -0.1in
\label{tab:ablation1}
\end{table*}
\begin{figure*}[t]
\begin{center}
\centerline{\includegraphics[width=1\columnwidth]{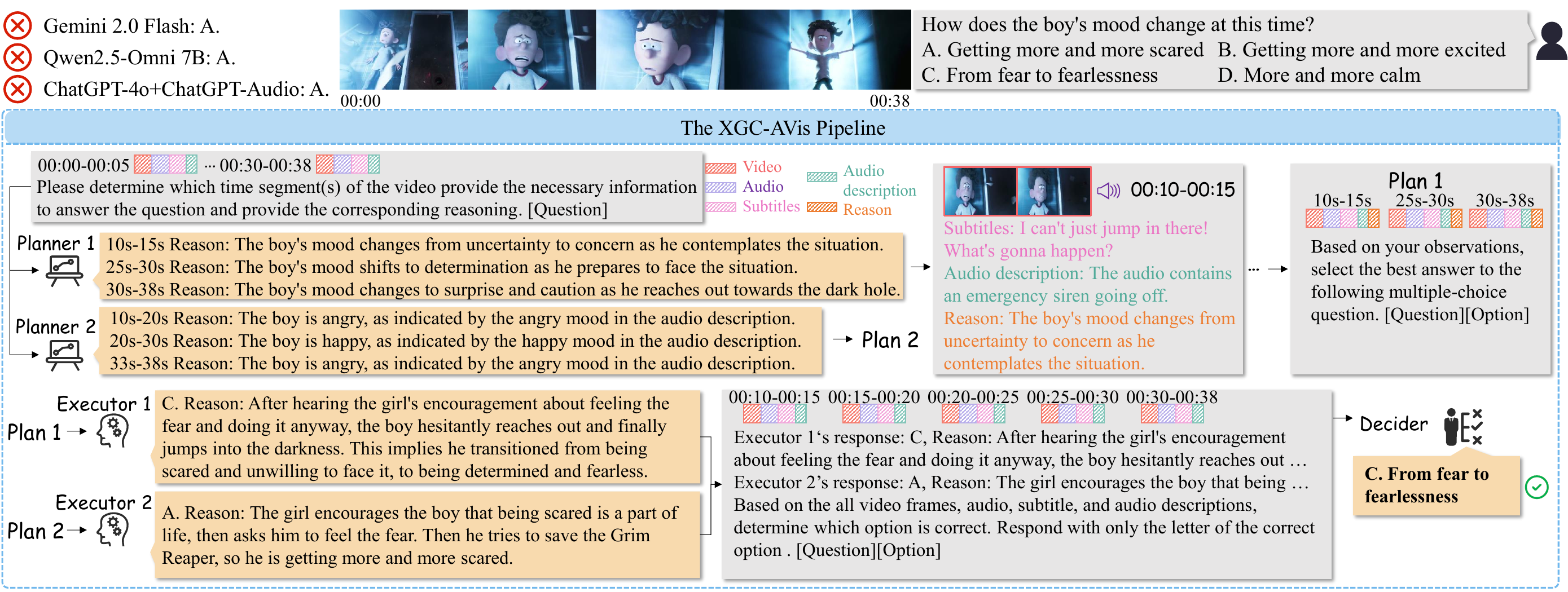}}
\vspace{-0.3em}
\caption{Example of the XGC-AVis response process.}
\label{fig:pipline}
\end{center}
\vspace{-2.5em}
\end{figure*}

\subsection{Ablation Study of XGC-AVis}
We evaluate the effectiveness of two planners in XGC-AVis on XGC-AVQuiz, with results shown in Table \ref{tab:ablation1}. Gemini 2.0 Flash serves as the baseline model. In XGC-AVis \textit{w/o} Planner 2, the final answer is directly taken from Executor 1, while in XGC-AVis \textit{w/o} Planner 1, it is taken from Executor 2. In both cases, removing either planner leads to a performance drop, demonstrating that the decider's ability to select the correct answer between Executor 1 and Executor 2 significantly improves accuracy and validates the effectiveness of the XGC-AVis architecture. Notably, the improvement is most pronounced in the A/V quality perception category, providing strong evidence that XGC-AVis effectively enhances MLLMs' performance on quality perception tasks.

\subsection{Case Study}
Fig. \ref{fig:pipline} presents a case study of how XGC-AVis answers an audio-visual question. In this example, the question is ``How does the boy's mood change at this time?'' XGC-AVis first concatenates the pre-processed video frames, audio segments, subtitles, and audio descriptions with the question, and feeds them into Planner 1 and Planner 2. Each planner identifies the relevant time segments and provides corresponding reasoning, as shown in their outputs. Based on Planner 1's selected segments, the system constructs Plan 1 and passes it to Executor 1, while Planner 2's outputs form Plan 2, which is processed by Executor 2. If the two executors produce inconsistent answers, XGC-AVis integrates the time spans and reasoning from both plans, along with the candidate answers, and forwards them to the Decider. The Decider then determines the final correct answer.

\section{Conclusion}
In this paper, we propose XGC-AVis, a multi-agent system that enhances MLLMs' audio-visual understanding capabilities without additional training. We also introduced XGC-AVQuiz, a novel benchmark designed to comprehensively evaluate MLLMs in both real-world and AI-generated scenarios, with a dedicated focus on quality perception. Experimental results reveal that MLLMs face significant challenges in quality perception and in temporally localizing audio-visual events. Ablation studies demonstrate that XGC-AVis can guide MLLMs to identify question-relevant segments, thereby enhancing temporal localization. Moreover, by integrating decisions from two executors through a decider, XGC-AVis further improves MLLMs' performance in the quality perception category.

\section*{Statments}

\subsection*{Ethics Statement}
This research adheres to the ICLR Code of Ethics. In developing the XGC-AVQuiz benchmark, we invited human annotators to label question-answer pairs. All participants were informed of the task's purpose and their participation was voluntary. No personally identifiable information was collected, and all data was anonymized to ensure privacy and confidentiality. We ensured that the annotation process was free of harm or discomfort for the participants. The annotations were collected solely for academic research, and no sensitive or inappropriate content was involved. Additionally, we took measures to avoid bias in the data collection and ensure fairness in the labeling process. This study complies with relevant ethical standards and legal requirements, with no conflicts of interest or financial sponsors influencing the research. We also ensured that the data usage and participant involvement adhered to ethical guidelines regarding privacy and research integrity.

\subsection*{Reproducibility Statement}
We have made significant efforts to ensure the reproducibility of our work. Detailed descriptions of the XGC-AVis framework can be found in Section \ref{sec:XGC-AVis} and Appendix \ref{sec:XGC-AVis Details}, while Section \ref{sec:XGC-AVQuiz} and Appendix \ref{sec:MLLM Experiment Details} provide the specifics of the XGC-AVQuiz benchmark. For the models and algorithms used in our experiments, we rely on open-source code and pretrained model weights, or publicly available official APIs, ensuring transparency and reproducibility. Additionally, after the final version of the paper is accepted, we will make the code and dataset publicly available to further facilitate reproducibility and transparency.

\subsection*{Additional LLM Statement}
We would like to clarify that LLMs were only used for language polishing and grammar correction in this work. The core research ideas, conceptual framework, experimental design, and data analysis are entirely the original work of the authors. The authors take full responsibility for the accuracy, completeness, and all statements made in this paper.
\bibliography{iclr2026_conference}

\begin{thebibliography}{42}
\providecommand{\natexlab}[1]{#1}
\providecommand{\url}[1]{\texttt{#1}}
\expandafter\ifx\csname urlstyle\endcsname\relax
  \providecommand{\doi}[1]{doi: #1}\else
  \providecommand{\doi}{doi: \begingroup \urlstyle{rm}\Url}\fi

\bibitem[kli(2024)]{kling}
Kling.
\newblock Accessed June 6, 2024 [Online] \url{https://klingai.kuaishou.com/}, 2024.
\newblock URL \url{https://klingai.kuaishou.com/}.

\bibitem[Anthropic(2024)]{claude3}
Anthropic.
\newblock Claude 3 technical report.
\newblock \url{https://www.anthropic.com/news/claude-3}, 2024.
\newblock Accessed: 2025-09-23.

\bibitem[Bai et~al.(2023)Bai, Bai, Chu, Cui, Dang, Deng, Fan, Ge, Han, Huang, et~al.]{bai2023qwen}
Jinze Bai, Shuai Bai, Yunfei Chu, Zeyu Cui, Kai Dang, Xiaodong Deng, Yang Fan, Wenbin Ge, Yu~Han, Fei Huang, et~al.
\newblock Qwen technical report.
\newblock \emph{arXiv preprint arXiv:2309.16609}, 2023.

\bibitem[Benchekroun et~al.(2023)Benchekroun, Dervishi, Ibrahim, Gaya, Martinet, Mialon, Scialom, Dupoux, Hupkes, and Vincent]{benchekroun2023worldsense}
Youssef Benchekroun, Megi Dervishi, Mark Ibrahim, Jean-Baptiste Gaya, Xavier Martinet, Gr{\'e}goire Mialon, Thomas Scialom, Emmanuel Dupoux, Dieuwke Hupkes, and Pascal Vincent.
\newblock Worldsense: A synthetic benchmark for grounded reasoning in large language models.
\newblock \emph{arXiv preprint arXiv:2311.15930}, 2023.

\bibitem[Cao et~al.(2023)Cao, Min, Sun, and Zhai]{cao2023subjective}
Yuqin Cao, Xiongkuo Min, Wei Sun, and Guangtao Zhai.
\newblock Subjective and objective audio-visual quality assessment for user generated content.
\newblock \emph{IEEE Transactions on Image Processing}, 32:\penalty0 3847--3861, 2023.

\bibitem[Cao et~al.(2025)Cao, Min, Gao, Sun, and Zhai]{cao2025agav}
Yuqin Cao, Xiongkuo Min, Yixuan Gao, Wei Sun, and Guangtao Zhai.
\newblock Agav-rater: adapting large multimodal model for ai-generated audio-visual quality assessment.
\newblock \emph{International Conference on Machine Learning}, 2025.

\bibitem[Chen et~al.(2022)Chen, Wu, Wang, Liu, Tompkins, Chen, and Wei]{chen2022beats}
Sanyuan Chen, Yu~Wu, Chengyi Wang, Shujie Liu, Daniel Tompkins, Zhuo Chen, and Furu Wei.
\newblock Beats: Audio pre-training with acoustic tokenizers.
\newblock \emph{arXiv preprint arXiv:2212.09058}, 2022.

\bibitem[Chen et~al.(2024)Chen, Wang, Cao, Liu, Gao, Cui, Zhu, Ye, Tian, Liu, et~al.]{chen2024expanding}
Zhe Chen, Weiyun Wang, Yue Cao, Yangzhou Liu, Zhangwei Gao, Erfei Cui, Jinguo Zhu, Shenglong Ye, Hao Tian, Zhaoyang Liu, et~al.
\newblock Expanding performance boundaries of open-source multimodal models with model, data, and test-time scaling.
\newblock \emph{arXiv preprint arXiv:2412.05271}, 2024.

\bibitem[Cheng et~al.(2024)Cheng, Leng, Zhang, Xin, Li, Chen, Zhu, Zhang, Luo, Zhao, et~al.]{cheng2024videollama}
Zesen Cheng, Sicong Leng, Hang Zhang, Yifei Xin, Xin Li, Guanzheng Chen, Yongxin Zhu, Wenqi Zhang, Ziyang Luo, Deli Zhao, et~al.
\newblock Videollama 2: Advancing spatial-temporal modeling and audio understanding in video-llms.
\newblock \emph{arXiv preprint arXiv:2406.07476}, 2024.

\bibitem[DeepMind(2025)]{gemini20flash}
Google DeepMind.
\newblock Gemini 2.0 flash.
\newblock \url{https://deepmind.google/technologies/gemini/}, 2025.
\newblock Accessed: 2025-09-23.

\bibitem[Dosovitskiy et~al.(2020)Dosovitskiy, Beyer, Kolesnikov, Weissenborn, Zhai, Unterthiner, Dehghani, Minderer, Heigold, Gelly, et~al.]{dosovitskiy2020image}
Alexey Dosovitskiy, Lucas Beyer, Alexander Kolesnikov, Dirk Weissenborn, Xiaohua Zhai, Thomas Unterthiner, Mostafa Dehghani, Matthias Minderer, Georg Heigold, Sylvain Gelly, et~al.
\newblock An image is worth 16x16 words: Transformers for image recognition at scale.
\newblock \emph{arXiv preprint arXiv:2010.11929}, 2020.

\bibitem[Gong et~al.(2024)Gong, Feng, Li, Wang, Cheng, Yang, Han, Wang, Bai, Yang, et~al.]{gong2024av}
Kaixiong Gong, Kaituo Feng, Bohao Li, Yibing Wang, Mofan Cheng, Shijia Yang, Jiaming Han, Benyou Wang, Yutong Bai, Zhuoran Yang, et~al.
\newblock Av-odyssey bench: Can your multimodal llms really understand audio-visual information?
\newblock \emph{arXiv preprint arXiv:2412.02611}, 2024.

\bibitem[Hsu et~al.(2021)Hsu, Bolte, Tsai, Lakhotia, Salakhutdinov, and Mohamed]{hsu2021hubert}
Wei-Ning Hsu, Benjamin Bolte, Yao-Hung~Hubert Tsai, Kushal Lakhotia, Ruslan Salakhutdinov, and Abdelrahman Mohamed.
\newblock Hubert: Self-supervised speech representation learning by masked prediction of hidden units.
\newblock \emph{IEEE/ACM transactions on audio, speech, and language processing}, 29:\penalty0 3451--3460, 2021.

\bibitem[Hurst et~al.(2024)Hurst, Lerer, Goucher, Perelman, Ramesh, Clark, Ostrow, Welihinda, Hayes, Radford, et~al.]{hurst2024gpt}
Aaron Hurst, Adam Lerer, Adam~P Goucher, Adam Perelman, Aditya Ramesh, Aidan Clark, AJ~Ostrow, Akila Welihinda, Alan Hayes, Alec Radford, et~al.
\newblock Gpt-4o system card.
\newblock \emph{arXiv preprint arXiv:2410.21276}, 2024.

\bibitem[Li et~al.(2024{\natexlab{a}})Li, Zhang, Guo, Zhang, Li, Zhang, Zhang, Zhang, Li, Liu, et~al.]{li2024llava}
Bo~Li, Yuanhan Zhang, Dong Guo, Renrui Zhang, Feng Li, Hao Zhang, Kaichen Zhang, Peiyuan Zhang, Yanwei Li, Ziwei Liu, et~al.
\newblock Llava-onevision: Easy visual task transfer.
\newblock \emph{arXiv preprint arXiv:2408.03326}, 2024{\natexlab{a}}.

\bibitem[Li et~al.(2023)Li, Zhang, Wu, Sun, Min, Liu, Zhai, and Lin]{li2023agiqa}
Chunyi Li, Zicheng Zhang, Haoning Wu, Wei Sun, Xiongkuo Min, Xiaohong Liu, Guangtao Zhai, and Weisi Lin.
\newblock Agiqa-3k: An open database for ai-generated image quality assessment.
\newblock \emph{IEEE Transactions on Circuits and Systems for Video Technology}, 34\penalty0 (8):\penalty0 6833--6846, 2023.

\bibitem[Li et~al.(2024{\natexlab{b}})Li, Liu, Wu, Wang, Shen, Qu, Niu, Zhou, Huang, Li, et~al.]{li2024aria}
Dongxu Li, Yudong Liu, Haoning Wu, Yue Wang, Zhiqi Shen, Bowen Qu, Xinyao Niu, Fan Zhou, Chengen Huang, Yanpeng Li, et~al.
\newblock Aria: An open multimodal native mixture-of-experts model.
\newblock \emph{arXiv preprint arXiv:2410.05993}, 2024{\natexlab{b}}.

\bibitem[Li et~al.(2025)Li, Liu, Dinkel, Niu, Zhang, and Luan]{li2025reinforcement}
Gang Li, Jizhong Liu, Heinrich Dinkel, Yadong Niu, Junbo Zhang, and Jian Luan.
\newblock Reinforcement learning outperforms supervised fine-tuning: A case study on audio question answering.
\newblock \emph{arXiv preprint arXiv:2503.11197}, 2025.

\bibitem[Li et~al.(2022)Li, Wei, Tian, Xu, Wen, and Hu]{li2022learning}
Guangyao Li, Yake Wei, Yapeng Tian, Chenliang Xu, Ji-Rong Wen, and Di~Hu.
\newblock Learning to answer questions in dynamic audio-visual scenarios.
\newblock In \emph{Proceedings of the IEEE/CVF conference on computer vision and pattern recognition}, pp.\  19108--19118, 2022.

\bibitem[Li et~al.(2024{\natexlab{c}})Li, Wang, He, Li, Wang, Liu, Wang, Xu, Chen, Luo, et~al.]{li2024mvbench}
Kunchang Li, Yali Wang, Yinan He, Yizhuo Li, Yi~Wang, Yi~Liu, Zun Wang, Jilan Xu, Guo Chen, Ping Luo, et~al.
\newblock Mvbench: A comprehensive multi-modal video understanding benchmark.
\newblock In \emph{Proceedings of the IEEE/CVF Conference on Computer Vision and Pattern Recognition}, pp.\  22195--22206, 2024{\natexlab{c}}.

\bibitem[Li et~al.(2024{\natexlab{d}})Li, Zhang, Ma, Yuan, Zhu, Guo, Liang, Liu, Wang, Yang, et~al.]{li2024omnibench}
Yizhi Li, Ge~Zhang, Yinghao Ma, Ruibin Yuan, Kang Zhu, Hangyu Guo, Yiming Liang, Jiaheng Liu, Zekun Wang, Jian Yang, et~al.
\newblock Omnibench: Towards the future of universal omni-language models.
\newblock \emph{arXiv preprint arXiv:2409.15272}, 2024{\natexlab{d}}.

\bibitem[Li et~al.(2024{\natexlab{e}})Li, Xu, Zhang, Song, Cai, Qi, Zhou, Pan, Li, Vu, et~al.]{li2024groundinggpt}
Zhaowei Li, Qi~Xu, Dong Zhang, Hang Song, Yiqing Cai, Qi~Qi, Ran Zhou, Junting Pan, Zefeng Li, Van~Tu Vu, et~al.
\newblock Groundinggpt: Language enhanced multi-modal grounding model.
\newblock \emph{arXiv preprint arXiv:2401.06071}, 2024{\natexlab{e}}.

\bibitem[Liu et~al.(2024)Liu, Feng, Xue, Wang, Wu, Lu, Zhao, Deng, Zhang, Ruan, et~al.]{liu2024deepseek}
Aixin Liu, Bei Feng, Bing Xue, Bingxuan Wang, Bochao Wu, Chengda Lu, Chenggang Zhao, Chengqi Deng, Chenyu Zhang, Chong Ruan, et~al.
\newblock Deepseek-v3 technical report.
\newblock \emph{arXiv preprint arXiv:2412.19437}, 2024.

\bibitem[Lu et~al.(2024)Lu, Clark, Lee, Zhang, Khosla, Marten, Hoiem, and Kembhavi]{lu2024unified}
Jiasen Lu, Christopher Clark, Sangho Lee, Zichen Zhang, Savya Khosla, Ryan Marten, Derek Hoiem, and Aniruddha Kembhavi.
\newblock Unified-io 2: Scaling autoregressive multimodal models with vision language audio and action.
\newblock In \emph{Proceedings of the IEEE/CVF Conference on Computer Vision and Pattern Recognition}, pp.\  26439--26455, 2024.

\bibitem[Radford et~al.(2021)Radford, Kim, Hallacy, Ramesh, Goh, Agarwal, Sastry, Askell, Mishkin, Clark, et~al.]{radford2021learning}
Alec Radford, Jong~Wook Kim, Chris Hallacy, Aditya Ramesh, Gabriel Goh, Sandhini Agarwal, Girish Sastry, Amanda Askell, Pamela Mishkin, Jack Clark, et~al.
\newblock Learning transferable visual models from natural language supervision.
\newblock In \emph{International conference on machine learning}, pp.\  8748--8763. PmLR, 2021.

\bibitem[Su et~al.(2023)Su, Lan, Li, Xu, Wang, and Cai]{su2023pandagpt}
Yixuan Su, Tian Lan, Huayang Li, Jialu Xu, Yan Wang, and Deng Cai.
\newblock Pandagpt: One model to instruction-follow them all.
\newblock \emph{arXiv preprint arXiv:2305.16355}, 2023.

\bibitem[Sun et~al.(2024)Sun, Yu, Tang, Chen, Tan, Li, Lu, Ma, Wang, and Zhang]{sun2024video}
Guangzhi Sun, Wenyi Yu, Changli Tang, Xianzhao Chen, Tian Tan, Wei Li, Lu~Lu, Zejun Ma, Yuxuan Wang, and Chao Zhang.
\newblock video-salmonn: Speech-enhanced audio-visual large language models.
\newblock \emph{arXiv preprint arXiv:2406.15704}, 2024.

\bibitem[Touvron et~al.(2023)Touvron, Lavril, Izacard, Martinet, Lachaux, Lacroix, Rozi{\`e}re, Goyal, Hambro, Azhar, et~al.]{touvron2023llama}
Hugo Touvron, Thibaut Lavril, Gautier Izacard, Xavier Martinet, Marie-Anne Lachaux, Timoth{\'e}e Lacroix, Baptiste Rozi{\`e}re, Naman Goyal, Eric Hambro, Faisal Azhar, et~al.
\newblock Llama: Open and efficient foundation language models.
\newblock \emph{arXiv preprint arXiv:2302.13971}, 2023.

\bibitem[Wang et~al.(2024)Wang, Zou, Lin, Sun, Liu, Zhang, Liu, Aw, and Chen]{wang2024audiobench}
Bin Wang, Xunlong Zou, Geyu Lin, Shuo Sun, Zhuohan Liu, Wenyu Zhang, Zhengyuan Liu, AiTi Aw, and Nancy~F Chen.
\newblock Audiobench: A universal benchmark for audio large language models.
\newblock \emph{arXiv preprint arXiv:2406.16020}, 2024.

\bibitem[Wang et~al.(2025{\natexlab{a}})Wang, Duan, Zhai, Wang, and Min]{wang2025aigv}
Jiarui Wang, Huiyu Duan, Guangtao Zhai, Juntong Wang, and Xiongkuo Min.
\newblock Aigv-assessor: benchmarking and evaluating the perceptual quality of text-to-video generation with lmm.
\newblock In \emph{Proceedings of the Computer Vision and Pattern Recognition Conference}, pp.\  18869--18880, 2025{\natexlab{a}}.

\bibitem[Wang et~al.(2025{\natexlab{b}})Wang, Duan, Zhao, Wang, Zhai, and Min]{wang2025lmm4lmm}
Jiarui Wang, Huiyu Duan, Yu~Zhao, Juntong Wang, Guangtao Zhai, and Xiongkuo Min.
\newblock Lmm4lmm: Benchmarking and evaluating large-multimodal image generation with lmms.
\newblock \emph{arXiv preprint arXiv:2504.08358}, 2025{\natexlab{b}}.

\bibitem[Wang et~al.(2025{\natexlab{c}})Wang, Xia, Zhu, and Xie]{wang2025u}
Ziqian Wang, Xianjun Xia, Xinfa Zhu, and Lei Xie.
\newblock U-sam: An audio language model for unified speech, audio, and music understanding.
\newblock \emph{arXiv preprint arXiv:2505.13880}, 2025{\natexlab{c}}.

\bibitem[Xu et~al.(2025)Xu, Guo, He, Hu, He, Bai, Chen, Wang, Fan, Dang, et~al.]{xu2025qwen2}
Jin Xu, Zhifang Guo, Jinzheng He, Hangrui Hu, Ting He, Shuai Bai, Keqin Chen, Jialin Wang, Yang Fan, Kai Dang, et~al.
\newblock Qwen2. 5-omni technical report.
\newblock \emph{arXiv preprint arXiv:2503.20215}, 2025.

\bibitem[Yang et~al.(2022)Yang, Wang, Duan, Chen, Hou, Jin, and Zhu]{yang2022avqa}
Pinci Yang, Xin Wang, Xuguang Duan, Hong Chen, Runze Hou, Cong Jin, and Wenwu Zhu.
\newblock Avqa: A dataset for audio-visual question answering on videos.
\newblock In \emph{Proceedings of the 30th ACM international conference on multimedia}, pp.\  3480--3491, 2022.

\bibitem[Ye et~al.(2024)Ye, Yu, Shao, Xie, Torr, and Cao]{ye2024cat}
Qilang Ye, Zitong Yu, Rui Shao, Xinyu Xie, Philip Torr, and Xiaochun Cao.
\newblock Cat: Enhancing multimodal large language model to answer questions in dynamic audio-visual scenarios.
\newblock In \emph{European Conference on Computer Vision}, pp.\  146--164, 2024.

\bibitem[Zhan et~al.(2024)Zhan, Dai, Ye, Zhou, Zhang, Liu, Zhang, Yuan, Zhang, Li, et~al.]{zhan2024anygpt}
Jun Zhan, Junqi Dai, Jiasheng Ye, Yunhua Zhou, Dong Zhang, Zhigeng Liu, Xin Zhang, Ruibin Yuan, Ge~Zhang, Linyang Li, et~al.
\newblock Anygpt: Unified multimodal llm with discrete sequence modeling.
\newblock \emph{arXiv preprint arXiv:2402.12226}, 2024.

\bibitem[Zhang et~al.()Zhang, Wu, Li, Zhou, Sun, Min, Chen, Liu, Lin, and Zhai]{zhangbench}
Zicheng Zhang, Haoning Wu, Chunyi Li, Yingjie Zhou, Wei Sun, Xiongkuo Min, Zijian Chen, Xiaohong Liu, Weisi Lin, and Guangtao Zhai.
\newblock A-bench: Are lmms masters at evaluating ai-generated images?
\newblock In \emph{The Thirteenth International Conference on Learning Representations}.

\bibitem[Zhang et~al.(2025)Zhang, Jia, Wu, Li, Chen, Zhou, Sun, Liu, Min, Lin, et~al.]{zhang2025q}
Zicheng Zhang, Ziheng Jia, Haoning Wu, Chunyi Li, Zijian Chen, Yingjie Zhou, Wei Sun, Xiaohong Liu, Xiongkuo Min, Weisi Lin, et~al.
\newblock Q-bench-video: Benchmark the video quality understanding of lmms.
\newblock In \emph{Proceedings of the Computer Vision and Pattern Recognition Conference}, pp.\  3229--3239, 2025.

\bibitem[Zhao et~al.(2023)Zhao, Lin, Zhou, Huang, Feng, and Kang]{zhao2023bubogpt}
Yang Zhao, Zhijie Lin, Daquan Zhou, Zilong Huang, Jiashi Feng, and Bingyi Kang.
\newblock Bubogpt: Enabling visual grounding in multi-modal llms.
\newblock \emph{arXiv preprint arXiv:2307.08581}, 2023.

\bibitem[Zhou et~al.(2025)Zhou, Wang, and Wu]{zhou2025daily}
Ziwei Zhou, Rui Wang, and Zuxuan Wu.
\newblock Daily-omni: Towards audio-visual reasoning with temporal alignment across modalities.
\newblock \emph{arXiv preprint arXiv:2505.17862}, 2025.

\bibitem[Zhu et~al.(2025{\natexlab{a}})Zhu, Wang, Chen, Liu, Ye, Gu, Tian, Duan, Su, Shao, et~al.]{zhu2025internvl3}
Jinguo Zhu, Weiyun Wang, Zhe Chen, Zhaoyang Liu, Shenglong Ye, Lixin Gu, Hao Tian, Yuchen Duan, Weijie Su, Jie Shao, et~al.
\newblock Internvl3: Exploring advanced training and test-time recipes for open-source multimodal models.
\newblock \emph{arXiv preprint arXiv:2504.10479}, 2025{\natexlab{a}}.

\bibitem[Zhu et~al.(2025{\natexlab{b}})Zhu, Duan, Zhang, Zhu, Zhu, Teng, Min, and Zhai]{zhu2025does}
Yuxin Zhu, Huiyu Duan, Kaiwei Zhang, Yucheng Zhu, Xilei Zhu, Long Teng, Xiongkuo Min, and Guangtao Zhai.
\newblock How does audio influence visual attention in omnidirectional videos? database and model.
\newblock \emph{IEEE Transactions on Image Processing}, 2025{\natexlab{b}}.

\end{thebibliography}
\bibliographystyle{iclr2026_conference}

\newpage
\appendix

\section{Appendix}
\subsection{MLLM Experiment Details}
\label{sec:MLLM Experiment Details}
In the XGC-AVQuiz benchmark, each question contains $2$ to $6$ carefully designed options, with only one correct answer. The correct and incorrect answers are shuffled during the evaluation process. During the evaluation of MLLMs, we uniformly select $15$ frames from each video and provide the complete audio segment as input. 

\textbf{Prompt Template for MLLM Evaluation}

\#\textit{User: [video tokens][audio tokens]}

\textit{These are the frames of the video and the corresponding audio. Please select the best answer to the following multiple-choice question based on the video. Respond with only the letter [the range of option letters (e.g., A, B, C, D).] of the correct option.}

\textit{Question: [Multiple-choice Questions and Options]}

All MLLMs are tested in a zero-shot setting. For open-source MLLMs, we conducted tests by downloading the official default parameters and running the tests on two A100 GPUs with 160 GB of memory. Closed-source MLLMs are evaluated via official APIs to ensure the reproducibility of the results. Apart from ChatGPT-4o, we used the official Deepgram API to convert the audio content of each video into text, which is then used as subtitles for testing MLLM performance. 

\textbf{Prompt Template for MLLM Evaluation with Subtitles}

\#\textit{User: [video tokens][audio tokens]}

\textit{These are the frames of the video and the corresponding audio. The subtitle of this video: [subtitle]. Please select the best answer to the following multiple-choice question based on the video. Respond with only the letter [the range of option letters (e.g., A, B, C, D).] of the correct option.}

For ChatGPT-4o, the official ChatGPT-Audio is used, and the description of the audio by ChatGPT-Audio is input as subtitles to assist ChatGPT-4o in understanding the audio-visual content.

\textbf{Prompt Template for ChatGPT-Audio}

\#\textit{User:Listen to the audio carefully. Describe the sounds in the audio, convert the spoken words into accurate subtitles. [audio tokens]}

\textit{Question: [Multiple-choice Questions and Options]}

\subsection{Details on LLM-assisted Evaluation}
\label{sec:Details on LLM-assisted Evaluation}
During the experiment, we found that some MLLMs models output answers in a format that does not meet the requirements. For example, the MLLMs would not respond with just the letter of the option, but would instead add additional explanatory words. To address this, we adopted an LLM-assisted evaluation method. We input the question, options, correct answer, and the MLLM's response into an LLM to evaluate the accuracy of the answer. We used Qwen-plus to assist in judging the accuracy of MLLM responses. To reduce the inherent variability of large language models, where identical prompts may produce uncertain responses, we employed a five-round voting strategy. For each question-answer pair, we sent the prompt defined in the template below five times and determined the correctness of the answer based on the majority, selecting the result that appeared three or more times. When the response from the LLM-assisted evaluation does not meet the requirements, we manually verify the accuracy of the MLLM's answer.

\textbf{Prompt Templates for LLM Evaluation of MLLMs' Responses.}

\#\textit{User: Given the question [multiple-choice question and options, the correct answer is the option [correct answer]. The respondent's answer is [MLLM's answer]. Determine if the respondent's answer is correct (1) or incorrect (0). If uncertain, also provide 0. Only return the result as a single digit.]}

\subsection{XGC-AVis Details}
\label{sec:XGC-AVis Details}
\textbf{Audio Perception.} For XGC-AVis, we first use the Deepgram API as a translator to convert the speech in the audio into subtitles. Then, we utilize r1-aqa as an audio descriptor to describe the audio in segments. When the Deepgram API returns subtitles, it also provides the timestamps of each sentence. Therefore, when Deepgram detects speech within an audio segment, we input the detected subtitle along with it into r1-aqa to assist in the speech description.

\textbf{Prompt Templates for Audio Descriptor with Speech}

\#\textit{User: The subtitle of this audio: [subtitle]. Please describe the background sounds and music. Do not respond using list or dictionary formats. Instead, write your response in full sentences.}

If Deepgram detects that the current audio segment does not contain subtitles, we will prompt r1-aqa that there is no clear speech in the current audio segment, thereby assisting r1-aqa in providing a better description of the audio.

\textbf{Prompt Templates for Audio Descriptor without Clear Speech}

\#\textit{User: This audio does not contain clear speech. Please the background sounds, emotion and music. Do not respond using list or dictionary formats. Instead, write your response in full sentences.}

\textbf{Plan Generation.} After concatenating the video segment, audio segment, subtitle, and audio description, we input them into Planner 1 and Planner 2 to obtain the time segments related to the question, with Aria as Planner 1 and Qwen 2.5-Omni as Planner 2. Aria can only process video and text, while Qwen 2.5-Omni can process video, audio, and text simultaneously.

\textbf{Prompt Templates for Planner}

\#\textit{User: [video tokens][audio tokens] The above frames and audio are extracted
from [Time range]. Subtitle: [Subtitle]. Audio description: [Audio description]...}

\textit{The video lasts for [duration] seconds. Please carefully read the questions related to audio-visual perception, understanding, and reasoning abilities, closely observe the video frames, audio, subtitle, and audio descriptions. Please determine which time segment(s) of the video provide the necessary information to answer the question and provide the corresponding reasoning. Question: [question] No need to answer the question itself, just identify the time range(s) in [start time] to [end time] and provide the corresponding reasoning.}

\textit{Reply me with a structured output in JSON format: "\{"time\_segments": [\{ "start\_time": , "end\_time": , "reasoning": \}...]\}" If there is no content related to the question, only answer 'No.'}

\textbf{Execution.} Based on the output of the Planner, we extract only the specified time segments and input them into Executor 1 and Executor 2, allowing them to determine the correct option and provide the corresponding reasoning. Both Executor 1 and Executor 2 are based on Gemini 2.0 Flash. If the Planner is unable to identify the video content related to the question, the entire video is assumed to be relevant to the question by default.

\textbf{Prompt Templates for Executor}

\#\textit{User: [video tokens][audio tokens] The above frames and audio are extracted
from [Time range]. Subtitle: [Subtitle]. Audio description: [Audio description]. Video description: [planner's reason]...}

\#\textit{Please closely observe the video frames, audio, subtitle, and audio descriptions. Based on your observations, select the best answer to the following multiple-choice question. Respond with only the letter [the range of option letters] of the correct option, followed by 'Reason:' and your reasoning. Question: [Multiple-choice Questions and Options]}

\textbf{Reflection.} If the answers from Executor 1 and Executor 2 are consistent, the answer is directly output. When the answers are inconsistent, the responses from Plan 1 and Plan 2 are merged and input into the decider, which will provide the final answer.

\#\textit{User: [video tokens][audio tokens] The above frames and audio are extracted
from [Time range]. Subtitle: [Subtitle]. Audio description: [Audio description]. Video description: [planner's reason]...}

\textit{Executor 1's response: [Answer 1], [Reason 1]. Executor 2's response: [Answer 2], [Reason 2]. For this multiple-choice question, there are two different answers. Based on the all video frames, audio, subtitle, and audio descriptions, determine which option is correct. Respond with only the letter [the range of option letters] of the correct option.}
\begin{figure*}[t]
\begin{center}
\centerline{\includegraphics[width=1\columnwidth]{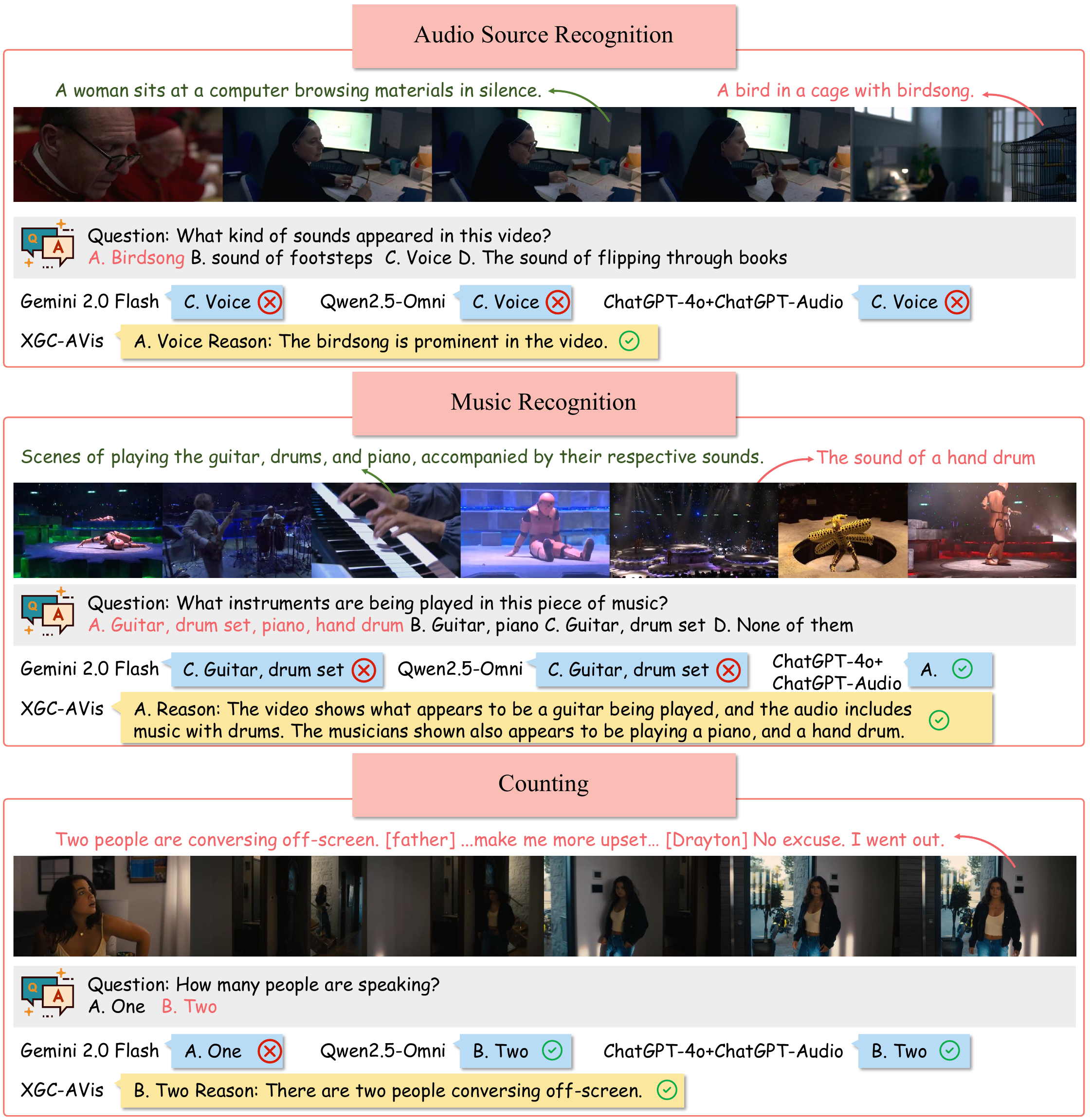}}
\caption{Examples of MLLM and AVis responses in the A/V recognition category, including audio source recognition, music recognition, and counting tasks.}
\label{fig:recognition}
\end{center}
\end{figure*}
\begin{figure*}[t]
\begin{center}
\centerline{\includegraphics[width=1\columnwidth]{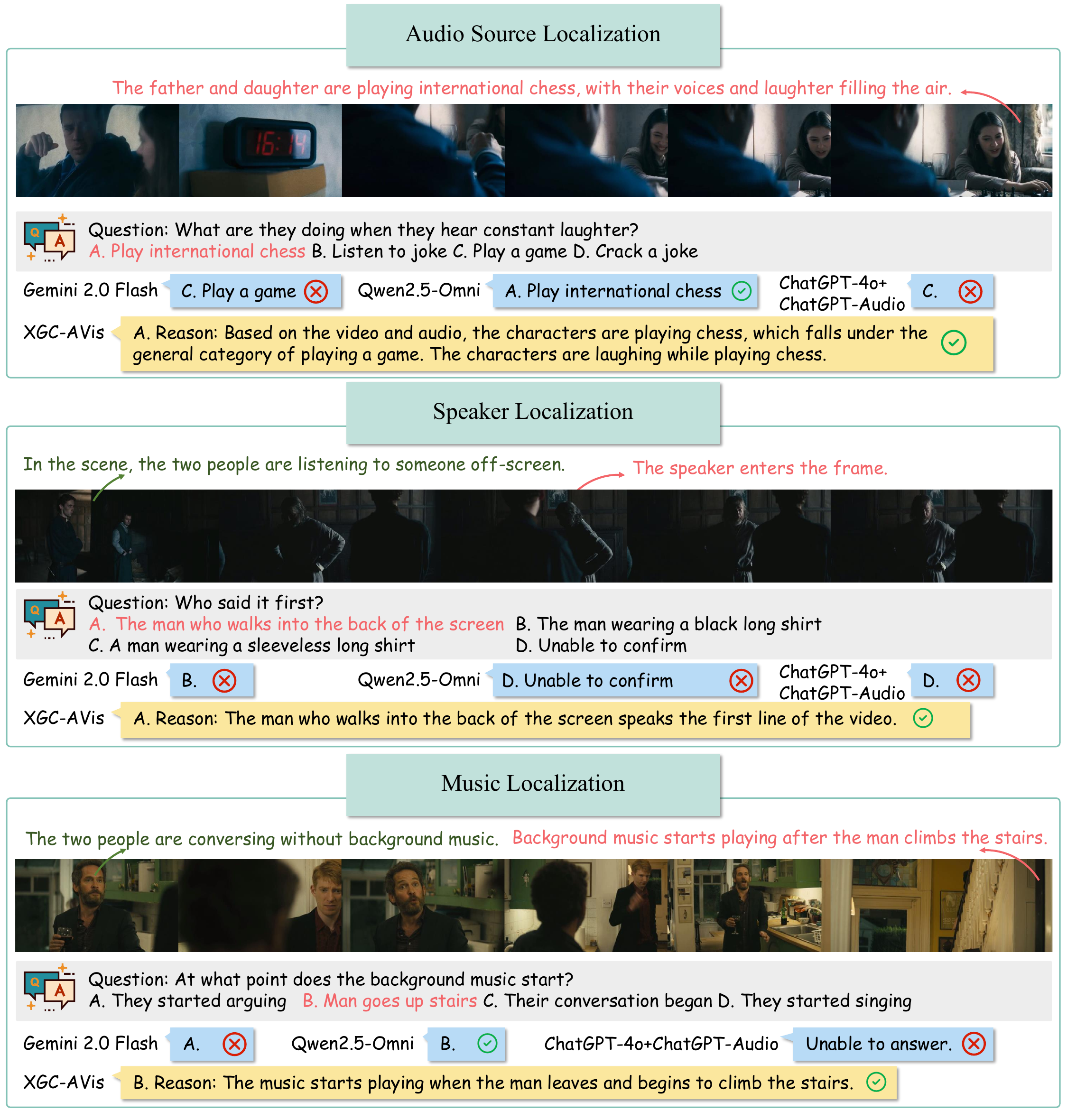}}
\caption{Examples of MLLM and XGC-AVis responses in the A/V localization category, including audio source localization, speaker localization and muisc localization.}
\label{fig:localization}
\end{center}
\end{figure*}
\subsection{QA Examples from Each Task}
\label{sec:QA Examples from Each Task}
XGC-AVQuiz encompasses $4$ A/V categories and $20$ A/V tasks. To provide a clearer understanding of each task type, we present one QA example for each task. 

\subsubsection{A/V Recognition Category}
Fig.~\ref{fig:recognition} illustrates $3$ representative tasks from the A/V recognition category: audio source recognition, music recognition, and counting.
The audio source recognition task evaluates whether MLLMs can correctly identify the types of sounds and their corresponding sources in a video. The music recognition task focuses on assessing MLLMs' ability to recognize the genre or composition of music within the video. The counting task challenges MLLMs to determine the number of occurrences of a specific sound event.

\subsubsection{A/V Localization Category}
Fig.~\ref{fig:localization} illustrates $3$ representative tasks from the A/V localization category: audio source localization, speaker localization, and music localization.
The audio source localization task evaluates MLLMs' ability to identify the position of non-speech, non-music sound sources in the video, as well as track variations in the sound source's volume over time. The speaker localization task tests MLLMs' capability to determine the positions and the temporal sequence of speakers in the video. The music localization task requires the model to detect the moments when music enters the scene and identify the time points when the music changes.
\newpage
\begin{figure}[t]
\begin{center}
\centerline{\includegraphics[width=1\columnwidth]{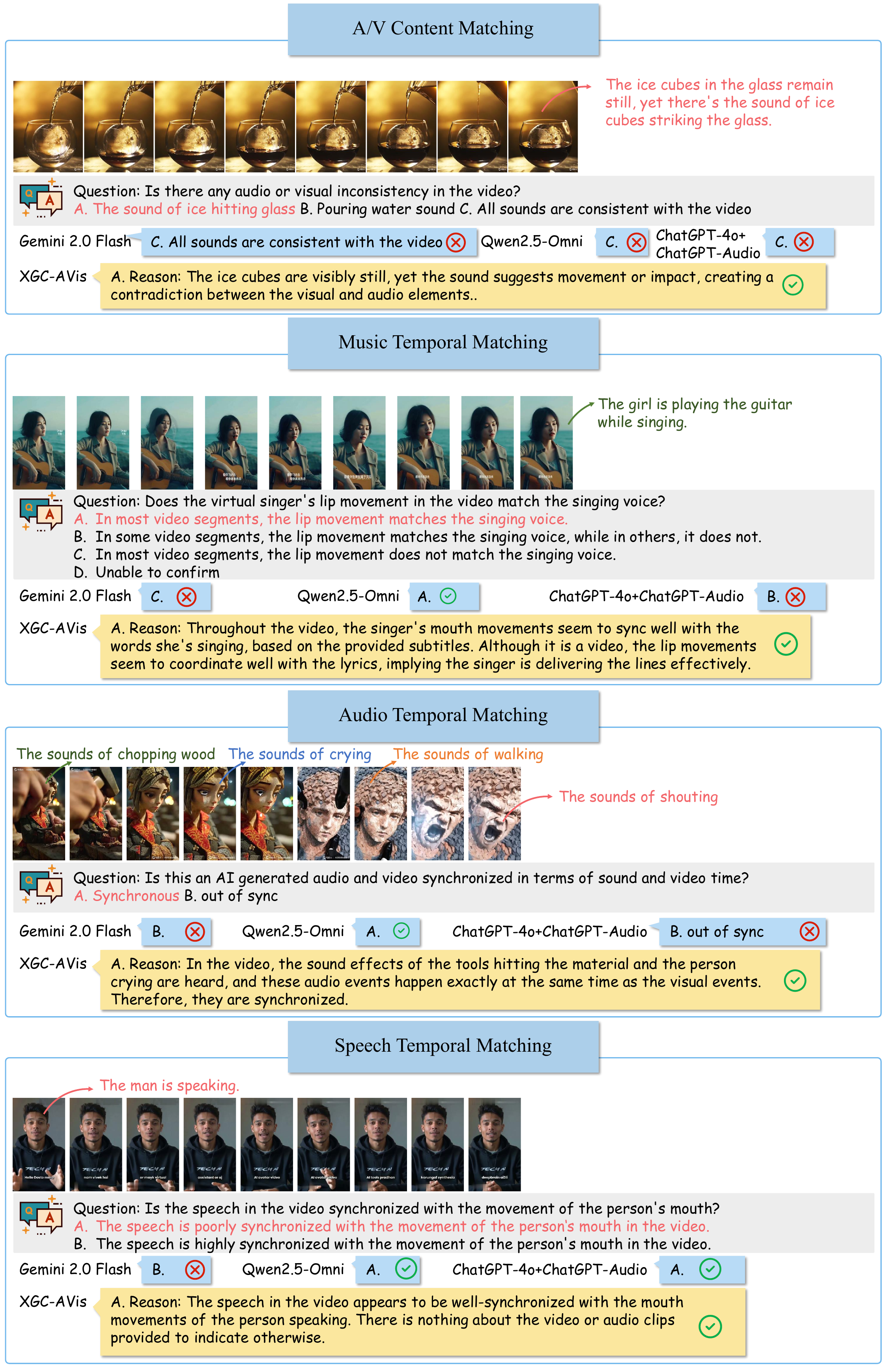}}
\caption{Examples of MLLM and XGC-AVis responses in the A/V quality perception category, including A/V content matching, music temporal matching, audio temporal matching and speech temporal matching.}
\label{fig:perception}
\end{center}
\end{figure}
\begin{figure*}[t]
\begin{center}
\centerline{\includegraphics[width=1\columnwidth]{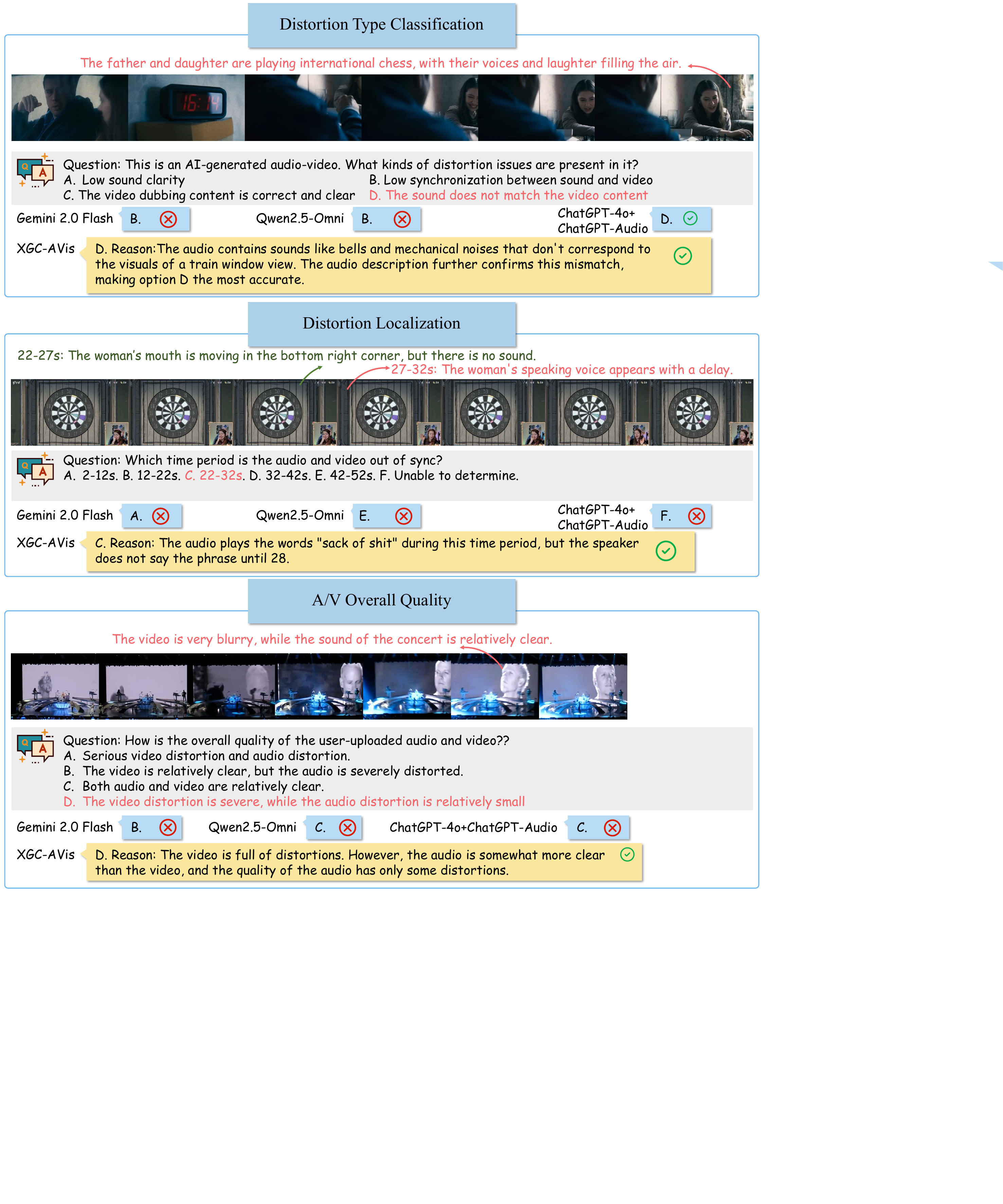}}
\caption{Examples of MLLM and XGC-AVis responses in the A/V quality perception category, including distortion type classification, distortion localization and A/V overall quality.}
\label{fig:perception1}
\end{center}
\end{figure*}
\subsubsection{A/V Quality Perception Category}
Fig.~\ref{fig:perception} and Fig.~\ref{fig:perception1} illustrates $7$ representative tasks from the A/V quality perception category: A/V content matching, music temporal matching, audio temporal matching, speech temporal matching, distortion type classification, distortion localization and A/V overall quality. 
The A/V content matching task tests whether MLLMs can determine the degree of alignment between the video and audio content in AIGC audio-visual content.
The music temporal matching task assesses whether MLLMs can perceive the synchronization between music events and corresponding visual content, such as whether the singer's lip movements align with the music.
The audio temporal matching task evaluates the synchronization of non-speech and non-music audio events with the video content in terms of timing.
The speech temporal matching task requires MLLMs to assess the temporal alignment between speech and video content, such as whether the speaker's lip movements correspond to the spoken words.
The distortion type classification task classifies the type of distortion present in a given audio-visual content, such as noise, blur, or compression artifacts.
The distortion localization task identifies the specific regions in the video where distortions occur, localizing them both spatially and temporally.
The A/V overall quality task evaluates the overall perceptual quality of the audio-visual content, combining both audio and video attributes into a single quality score.

\begin{figure*}[!h]
\begin{center}
\centerline{\includegraphics[width=1\columnwidth]{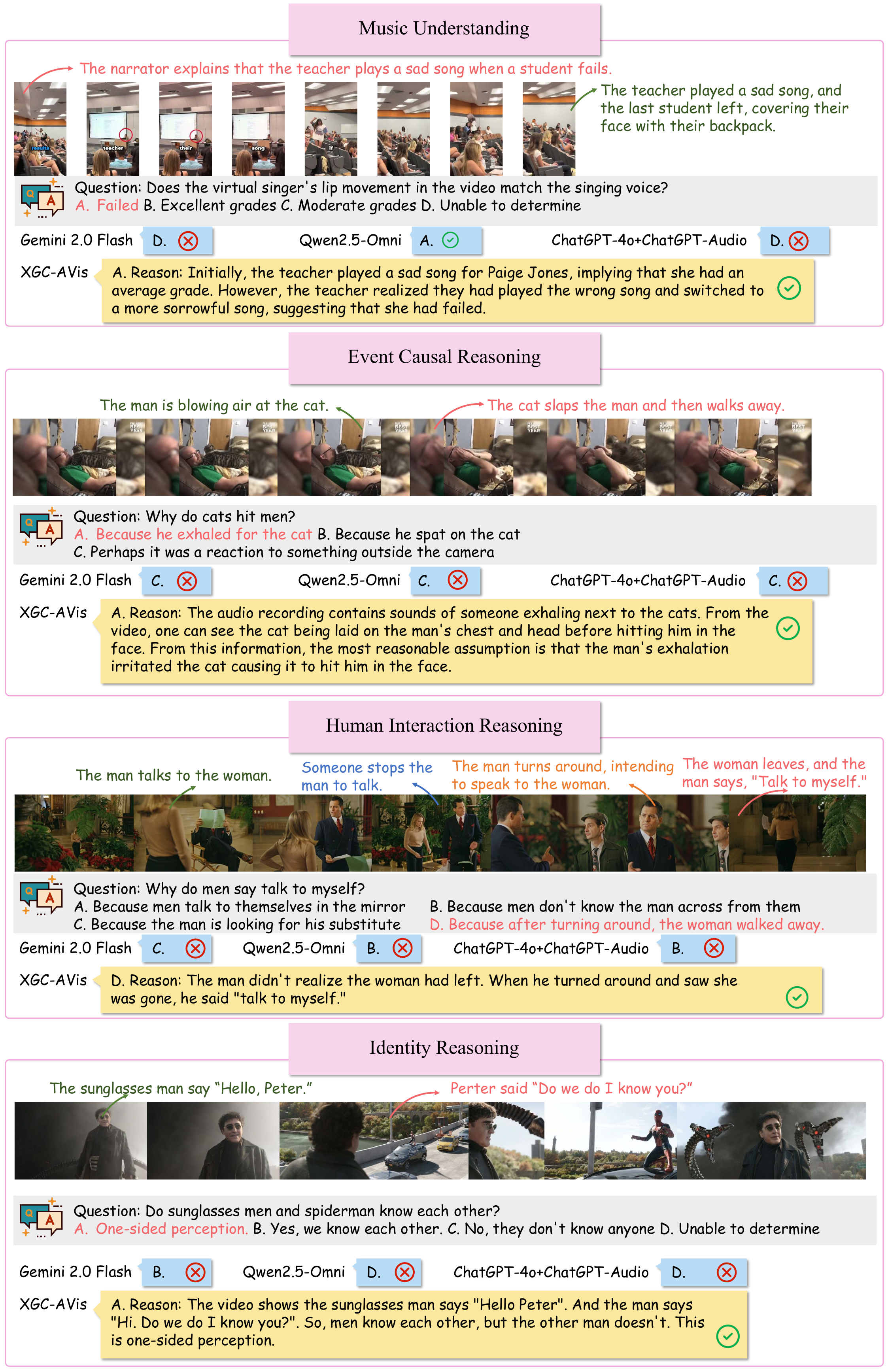}}
\caption{Examples of MLLM and XGC-AVis responses in the A/V reasoning category, including music understanding, event causal reasoning, human interaction reasoning and identity reasoning.}
\label{fig:reasoning}
\end{center}
\end{figure*}
\begin{figure*}[!h]
\begin{center}
\centerline{\includegraphics[width=1\columnwidth]{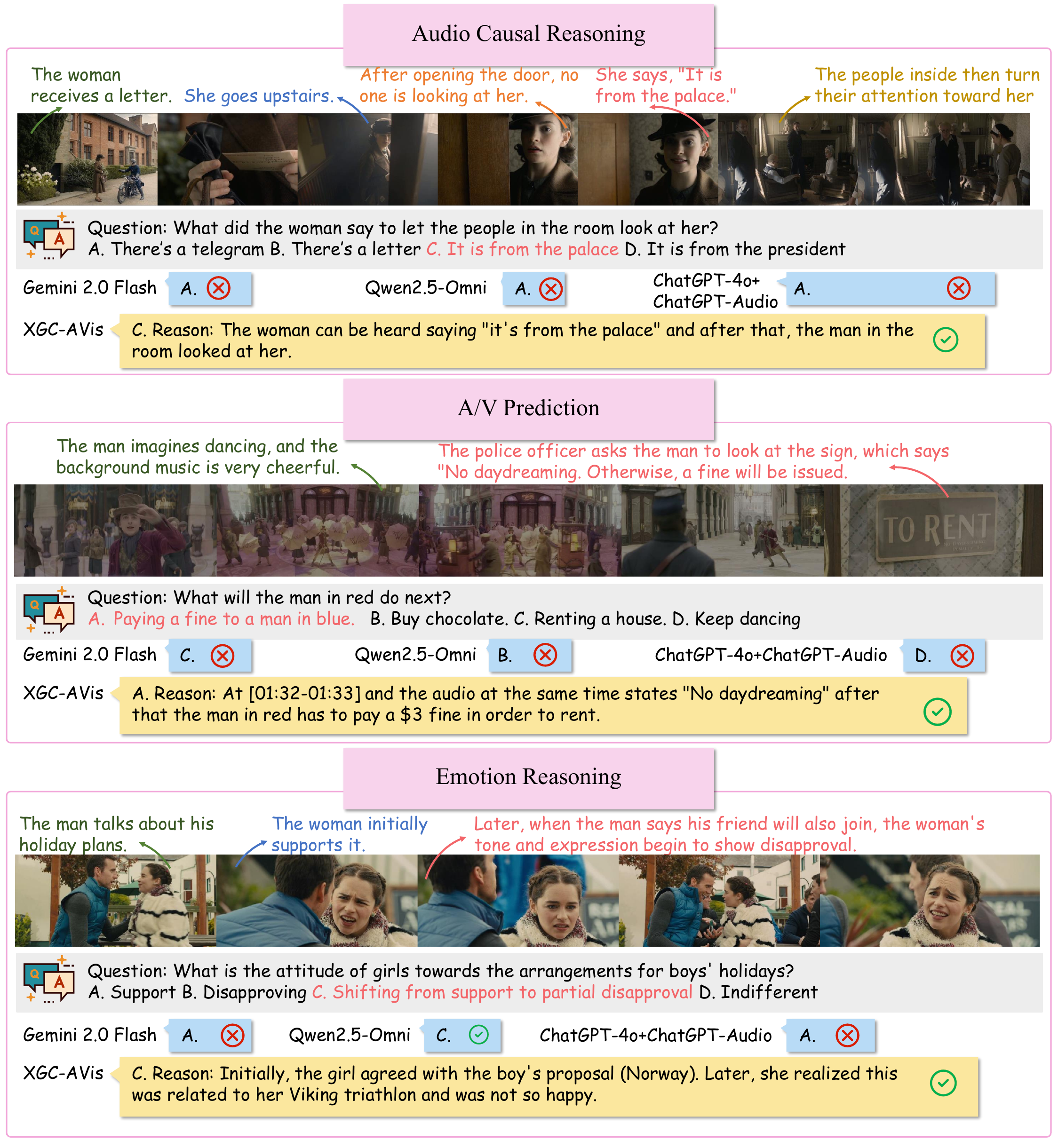}}
\caption{Examples of MLLM and XGC-AVis responses in the A/V reasoning category, including audio causal reasoning, A/V prediction, and emotion reasoning.}
\label{fig:reasoning1}
\end{center}
\end{figure*}
\subsubsection{A/V Reasoning Category}
Fig.~\ref{fig:reasoning} and Fig.~\ref{fig:reasoning1} illustrates $7$ representative tasks from the A/V reasoning category: music understanding, event causal reasoning, human interaction reasoning, identity reasoning, audio causal reasoning, A/V prediction, and emotion reasoning. 
The music understanding task requires MLLMs to comprehend the meaning conveyed by the music in the video or determine what aspects of the video the music aims to highlight.
The event causal reasoning task challenges MLLMs to reason about the storyline or the causes behind specific events based on the audio and video cues.
The human interaction reasoning task asks MLLMs to infer the reasons behind the actions of characters in the video, combining both audio and video events.
The identity reasoning task requires MLLMs to deduce the identity of characters or the relationships between two characters based on the audio-visual information.
The audio causal reasoning task directs MLLMs to focus on audio events, inferring the causes and effects of specific sounds within the video.
The A/V prediction task requires MLLMs to predict future events, behaviors, or sounds in the video by combining audio and visual information.
The emotion reasoning task involves judging emotions, tracking emotional changes, and understanding the causes behind these emotional shifts in the video content.
\end{document}